\documentclass[11pt,a4paper]{article}
\usepackage{jheppub}
\usepackage{orcidlink}
\usepackage{feynmp-auto}
\usepackage[utf8]{inputenc}
\usepackage[fleqn]{mathtools}
\usepackage[fleqn]{amsmath}
\usepackage{amssymb,mathrsfs}
\usepackage{setspace}
\usepackage{graphicx,floatflt,rotate, cancel}
\usepackage{bm}
\usepackage{color}
\usepackage{physics}
\usepackage{xcolor}
\usepackage{graphicx}
\usepackage{multirow}
\usepackage{array}
\usepackage{float}
\usepackage[toc,page]{appendix}
\usepackage{diagbox}
\usepackage[mathscr]{euscript}
\usepackage{cleveref}
\usepackage{soul}
\usepackage{pifont}
\usepackage{subfig}
\usepackage[numbers]{natbib}
\usepackage{notoccite}

\newcommand{\GeV}{\rm GeV}

\renewcommand\[{\begin{equation}}
\renewcommand\]{\end{equation}}

\makeatletter
\gdef\@fpheader{}
\makeatother

\definecolor{orange}{rgb}{1,0.5,0}

\title{Imprints of Left-Right Symmetry breaking in 
Gravitational Wave Spectroscopy:  A non-minimal case study }
\author[a]{Abdul Rahaman Shaikh \orcidlink{0000-0003-2295-027X}}
\emailAdd{arshaikh3921@gmail.com}
\affiliation[a]{Centre for Theoretical Physics, Jamia Millia Islamia, Jamia Nagar, New Delhi - 110025, India\looseness=-1}

\author[b,c]{, Nandini Das \orcidlink{0009-0003-6155-6205}}
\emailAdd{nandinidas.rs@gmail.com}
\affiliation[b]{SGTB Khalsa College, University of Delhi, Delhi, India-110007.}
\affiliation[c]{Department of Physics and Astrophysics, University of Delhi, Delhi, India-110007.}

\author[a]{, Rathin Adhikari \orcidlink{0000-0002-8764-9587}}
\emailAdd{rathin@ctp-jamia.res.in}

\author[d]{, Anindya Datta \orcidlink{0000-0002-4547-992X}}
\emailAdd{adphys@caluniv.ac.in}
\affiliation[d]{Department of Physics, University of Calcutta, 92, Acharya Prafulla Chandra Road, Kolkata 700009, India\looseness=-1}

\abstract{
We investigate the dynamics of cosmological phase transition in the framework of a non-minimal version of  Left-Right symmetric gauge model. At a high energy scale, this framework having the symmetry $SU(2)_L \otimes SU(2)_{R} \otimes U(1)_R \otimes U(1)_L$, is stemmed out from the Grand Unifying group $E_6$. Rich scalar sector of this model offers interesting opportunity for studying first order phase transition which can be the source of stochastic gravitational waves. Taking into account the theoretical and experimental constraints applicable to this model, we look for benchmark parameters where first order phase transition could be observed during the breaking of $SU(2)_{R} \otimes U(1)_R \otimes U(1)_L$ down to $U(1)_Y$, at an energy scale of tens of TeVs or higher.  Our  investigation reveals the possibility of strong first order phase transition over a large region of parameter space. We also discuss the production of stochastic gravitational waves arising from such strong first order phase transition. Our results present clear, testable predictions that fall directly within the sensitivity bands of ongoing and next-generation gravitational-wave observatories like LISA, DECIGO, AEDGE and $\mu$ARES.  
}

\begin{document}
\maketitle
\section{Introduction} \label{sec:into}
The detection of gravitational waves (GW) by the LIGO experiment \cite{LIGOScientific:2016aoc,PhysRevLett.116.241103,PhysRevLett.118.221101,LIGOScientific:2017vox,LIGOScientific:2017ycc} has opened up a window to look for the physics beyond the Standard Model (SM). The GW probe is novel in the sense that it helps us to probe into the nature of fundamental interactions at a very early epoch of the Universe when the average energy of the particles was much higher than the highest energy colliders available at present. At such high energies, the interactions among the fundamental particles, could have been governed by symmetries higher than the SM gauge symmetry, $SU(3)_C \otimes SU(2)_L \otimes U(1)_Y$ and probably plethora of other heavy particles  existed in the thermal bath along with the SM particles. As the Universe expands and its average temperature decreases, symmetry of the Universe would possibly reduced to the symmetry  of the SM in one or more steps. During such epochs of symmetry breaking, the Universe transits from a more symmetric phase to a less symmetric one. If the nature of such phase transition is of the first order, such a phenomena could be the origin of stochastic gravitational waves signals for which are being looked for at ground based experiments like LIGO \cite{LIGOScientific:2007fwp}, Virgo \cite{VIRGO:2012dcp} and KAGRA \cite{Aso:2013eba}; space bound experiment like LISA \cite{LISA:2017pwj} and pulsar timing array (PTA) like IIPTA \cite{2016MNRAS.458.1267V}. Recent observations by NANOGrav~\cite{NANOGrav:2023gor}, EPTA/InPTA~\cite{EPTA:2023fyk}, PPTA~\cite{Reardon:2023gzh}, and CPTA~\cite{Xu:2023wog} have provided evidence for a common-spectrum stochastic gravitational-wave background (SGWB) in the nano-hertz frequency range. While the signal is consistent with a source  from a population of supermassive black-hole binaries, a primordial cosmological origin cannot be comopletely excluded~\cite{NANOGrav:2023hvm}. In particular, gravitational waves generated by a cosmological first-order phase transition (FOPT) may provide a possible explanation of the observed background for such a GW signal~\cite{NANOGrav:2023hvm}. 

In the realm of microscopic distance scales and high energies, SM has emerged as an extraordinarily successful framework for describing the fundamental interactions among the elementary particles. The discovery of the Higgs boson at the Large Hadron Collider (LHC) in 2012~\cite{ATLAS:2012yve,CMS:2012qbp} and further efforts of precision measurements of the properties of the Higgs boson  at high-energy colliders have consistently validated the predictions of the SM. Despite its extraordinary success in describing elementary particle interactions, the SM cannot account for several well-established experimental and cosmological observations. Notable examples include the existence of dark matter (DM), the origin of neutrino masses and oscillations, and the observed matter-antimatter asymmetry of the Universe. These unresolved issues strongly indicate the presence of physics beyond the SM (BSM).

 Within the SM framework, two possible phase transitions may take place, the electroweak phase transition (EWPT) and the QCD chiral phase transition. However, non-perturbative lattice studies \cite{Csikor:1998eu} have established that, for the observed Higgs boson mass, the EWPT is a smooth crossover rather than a  FOPT, while the QCD transition is also a crossover for physical quark masses. Consequently, neither of the aforementioned transitions is expected to generate any interesting observable  signatures of primordial GW.  
 
 In contrast, many extensions of the SM with scalar potential constructed from multitude of scalar fields, may naturally accommodate FOPT. Non-trivial structure of the potential  provides more than one vacuum
 with different depths of which the deepest would be the true ground state of the system. If sufficiently strong, transitions between one of such false vacuum to the true (deepest) vacuum can result into sufficiently strong first order phase transition (SFOPT) which in turn  
 produce stochastic gravitational waves through bubble nucleation, expansion, and collision, as well as through sound waves and magnetohydrodynamic turbulence in the primordial plasma. These signals may be detectable by present and future gravitational-wave observatories like LISA~\cite{LISA:2017pwj}, AEDGE~\cite{AEDGE:2019nxb}, $\mu$ARES~\cite{Sesana:2019vho} etc. In the following, we would like to investigate a well motivated scenario of fundamental interactions, in which a first order phase transition 
 may take place due to its extended scalar sector.

In this work,  we focus on a variant  of left-right (LR) symmetric framework stemmed out from the grand unified symmetry group $E_6$~\cite{Shafi:1978gg,Gursey:1975ki}. $E_6$ symmetry first breaks to the gauge group $[SU(3)]^3$, which subsequently breaks to the LR gauge symmetry $SU(3)_C \otimes SU(2)_L \otimes SU(2)_R \otimes U(1)_L \otimes U(1)_R$, commonly referred to as the 32121 model. The collider phenomenology of this model has been investigated extensively in previous studies~\cite{Bhattacharyya:2021lgr}. However, in contrast to the conventional left-right symmetric model (LRSM) based on the $SU(3)_C \otimes SU(2)_L \otimes SU(2)_R \otimes U(1)_{B-L}$ gauge symmetry~\cite{Brdar:2019fur,Graf:2021xku}, the thermal history and cosmological evolution of the 32121 model remain unexplored. The primary goal of this work is to investigate the thermal evolution and phase transition dynamics of the 32121 model. To the best of our knowledge, a comprehensive study of the symmetry-breaking pattern and finite-temperature phase structure of this framework has not yet been carried out. Such an analysis is essential for understanding the cosmological implications of the model.

It is well known that gauge theories with multiple scalars (an example like ours) could be problematic due to infrared Linde problem \cite{LINDE1980289}.  At high temperatures, particularly near the critical temperature, some of the couplings become non-perturbative and standard techniques of  perturbative approach fail. However, in this work, we will study the first order phase transition in the context of perturbative LRSM framework.

One aspect of the present work definitely needs to highlighted. In the 32121 framework considered in the following, in principle, two phase transitions can take place and both of these could be of the first order in nature. The first one corresponds to the breaking of the LR gauge symmetry to the SM gauge group, while the second one, taking place at much lower energies, is the EWPT, which may also become first order due to the presence of the extended scalar sector. Study of EWPT in the framework of LRSMs has been  received considerable attention in the existing literature \cite{Choi:1992wb,Barenboim:1998ib, Dasgupta:2025uzi,Karmakar:2023ixo,Borboruah:2022eex}.  A couple of studies also looks into the LR symmetry breaking~\cite{Brdar:2019fur,Graf:2021xku} and among them only one~\cite{Borboruah:2022eex} studies two field phase transition in such a context. Our present study intends to investigate the possibility of a cosmological phase transition at the energy scale of LR symmetry breaking  in a two field configuration and resulting GW spectra. In that sense, the present study is somehow complementary to the previous studies in this direction.

The paper is organized as follows. In Section ~\ref{sec:model}, we present the 32121 model and discuss its essential features. Section~\ref{sec:constraints} is devoted to the theoretical and phenomenological constraints on the model parameter space. In Section~\ref{sec:effective_potential}, we derive the one-loop finite-temperature effective potential. The phase-transition analysis and the relevant thermodynamic quantities are discussed in Section~\ref{sec:phase_transition}. In Section~\ref{sec:na}, we present the results of our numerical analysis. In The GW signatures of the model and their detectability by current and future GW experiments will be discussed in Section~\ref{sec:GW}. Finally, we summarise in Section~\ref{sec:con}.
\section{Model} \label{sec:model}
In this work, we consider a LR extension of the SM. Similar to other LRSMs, this model can be viewed as an intermediate stage in the symmetry-breaking chain of a Grand Unified Theory. In our case, the model originates from the grand unified gauge group \(E_6\)~\cite{Ross:1985ai}.

Since our primary interest lies in the TeV-scale phenomenology of the model, we do not discuss the symmetry-breaking sequence or phase transitions leading from the \(E_6\) gauge group to the  \(32121\) gauge structure. Instead, we focus on the subsequent symmetry-breaking transition from the LRSM gauge group to the SM gauge group. In particular, we investigate whether this phase transition can be strongly first order and explore the resulting GW signatures that may be detectable by current and future ground- and space-based GW observatories.

In the 32121 model, the matter and gauge fields, along with their corresponding charges under the different gauge groups, are listed in Table~\ref{tab:fields}. Since any viable extension of the SM must ultimately reproduce the electromagnetic gauge symmetry \(U(1)_{\rm em}\), the electric charge operator \(Q\) in this model is related to the generators of \(SU(2)_L\), \(SU(2)_R\), \(U(1)_L\), and \(U(1)_R\) through
\(Q = T_{3L} + T_{3R} + \frac{Y_L}{2} + \frac{Y_R}{2}\) to ensures the correct embedding of the electromagnetic gauge symmetry after spontaneous symmetry breaking.
\begin{table}[h!]
	\centering
		\begin{tabular}{|c|c|c|c|c|c|c|}
			\hline
			&  & $SU(3)_C$ & $SU(2)_L$ & $SU(2)_R$ & $U(1)_L$ & $U(1)_R$ \\
			\hline 
			& $L_L$ & $1$ & $2$ & $1$ & $-1/6$ & $-1/3$  \\ 
			& $\bar{L}_R$ & $1$ & $1$ & $2$ & $1/3$ & $1/6$  \\ 
			& $\bar{L}_B$ & $1$ & $2$ & $2$ & $-1/6$ & $1/6$  \\ 
			Fermions & $\bar{L}_S$ & $1$ & $1$ & $1$ & $1/3$ & $-1/3$ \\ 
			& $Q_L$ & $3$ & $2$ & $1$ & $1/6$ & $0$ \\ 
			& $\bar{Q}_R$ & $\bar{3}$ & $1$ & $2$ & $0$ & $-1/6$ \\
			& $\bar{Q}_{LS}$ & $\bar{3}$ & $1$ & $1$ & $-1/3$ & $0$ \\ 
			& $Q_{RS}$ & $3$ & $1$ & $1$ & $0$ & $1/3$ \\
			\hline					
			& $\Phi_B$ & $1$ & $2$ & $2$ & $1/6$ & $-1/6$  \\ 
			Bosons & $\Phi_L$ & $1$ & $2$ & $1$ & $1/6$ & $1/3$  \\ 
			& $\Phi_R$ & $1$ & $1$ & $2$ & $-1/3$ & $-1/6$  \\ 
			& $\Phi_S$ & $1$ & $1$ & $1$ & $-1/3$ & $1/3$ \\ 
			\hline				
			& $G^i ,\; i=1,...,8$& $8$ & $1$ & $1$ & $0$ & $0$ \\
			& $W^i_{L}, i=1,2,3$ & $1$ & $3$ & $1$ & $0$ & $0$ \\ 
			Gauge bosons & $W^i_{R}, i=1,2,3$ & $1$ & $1$ & $3$ & $0$ & $0$  \\ 
			& $B_L$ & $1$ & $1$ & $1$ & $0$ & $0$ \\ 
			& $B_R$ & $1$ & $1$ & $1$ & $0$ & $0$ \\ 
			\hline
		\end{tabular}
	\caption{Fermions and Bosons and their respective quantum numbers in $32121$ model.}
\label{tab:fields}
\end{table}	

The $SU(2)_L$ of $32121$ is identified to the weak isospin group of the SM and $U(1)_Y$ of the SM will arise due to breaking of $SU(2)_{R} \otimes U(1)_L  \otimes U(1)_R$. So, the gauge coupling constant will obey the following relation:
\begin{equation}
	\dfrac{1}{g_Y^2} = \dfrac{1}{g_{2R}^2} + \dfrac{1}{g_{1L}^2} + \dfrac{1}{g_{1R}^2}.
    \label{eq:gauge_couplings_relation}
\end{equation}
Furthermore, we assume \(g_{2L}=g_{2R}\equiv g_2\) and \(g_{1L}=g_{1R}\equiv g_1\) in order to preserve the LR symmetry of the Lagrangian. This assumption is adopted throughout the remainder of this article and is used in all subsequent analyses.

The scalar sector of the model contains one Higgs bi-doublet ($\Phi_B$), one left-handed ($\Phi_L$), one right-handed ($\Phi_R$) weak doublets and a singlet Higgs boson ($\Phi_S$) with non-zero $U(1)$ charges. After all the scalars acquire $vev$, they can be written as
\begin{equation}
\begin{gathered}
\Phi_B=
\begin{pmatrix}
\dfrac{1}{\sqrt{2}}\left(v_1+\phi_1^0+i\xi_1^0\right) & \phi_1^+ \\
\phi_2^- & \dfrac{1}{\sqrt{2}}\left(v_2+\phi_2^0+i\xi_2^0\right)
\end{pmatrix},
\qquad
\Phi_S=
\dfrac{1}{\sqrt{2}}
\left(v_S+\phi_S^0+i\xi_S^0\right), \\[3mm]
\Phi_L=
\begin{pmatrix}
\phi_L^+ \\
\dfrac{1}{\sqrt{2}}\left(v_L+\phi_L^0+i\xi_L^0\right)
\end{pmatrix},
\qquad
\Phi_R=
\begin{pmatrix}
\dfrac{1}{\sqrt{2}}\left(v_R+\phi_R^0+i\xi_R^0\right) \\
\phi_R^-
\end{pmatrix}.
\end{gathered}
\label{eq:scalars}
\end{equation}
The most general scalar potential of our model can be  written as sum of two parts ${\cal V}_1$ and ${\cal V}_2$ given by
\begin{eqnarray}
	\mathcal{V}_1 = &-&\mu_1^2 Tr \left( {\Phi_B}^{\dagger} \Phi_B\right) - \mu_3^2 \left( {\Phi_L}^{\dagger} \Phi_L + {\Phi_R}^{\dagger} \Phi_R \right)  - \mu_4^2 {\Phi_S}^{\dagger} \Phi_S   \nonumber \\
	&+& \lambda_1 Tr \left[ ({\Phi_B}^{\dagger} \Phi_B)\right]^2 + \lambda_3 \left( Tr\left[ {\Phi_B}^{\dagger} \tilde{\Phi}_B\right]  Tr\left[ \tilde{\Phi}_B^{\dagger} \Phi_B\right] \right) \nonumber \\
	&+& \alpha_1 (\Phi_S^{\dagger} \Phi_S)^2 + \beta_1 Tr\left[ {\Phi_B}^{\dagger} \Phi_B\right]  (\Phi_S^{\dagger} \Phi_S) + \gamma_1 \left[ (\Phi_L^{\dagger} \Phi_L) + (\Phi_R^{\dagger} \Phi_R)\right] (\Phi_S^{\dagger} \Phi_S) \nonumber \\
	&+& \rho_1 \left[ (\Phi_L^{\dagger} \Phi_L)^2 + (\Phi_R^{\dagger} \Phi_R)^2\right] + \rho_3 \left[ (\Phi_L^{\dagger} \Phi_L) (\Phi_R^{\dagger} \Phi_R)\right] + c_1 Tr\left[ {\Phi_B}^{\dagger} \Phi_B\right] \left[ (\Phi_L^{\dagger} \Phi_L) + (\Phi_R^{\dagger} \Phi_R)\right]   \nonumber \\
	&+& c_3 \left[  ( \Phi_L^{\dagger} \Phi_B  \Phi_B^{\dagger} \Phi_L ) + ( \Phi_R^{\dagger} \Phi_B^{\dagger} \Phi_B  \Phi_R ) \right] + c_4 \left[ ( \Phi_L^{\dagger} \tilde{\Phi}_B \tilde{\Phi}_B^{\dagger} \Phi_L ) + ( \Phi_R^{\dagger} \tilde{\Phi}_B^{\dagger} \tilde{\Phi}_B \Phi_R ) \right]
\label{eq:pot_zero}
\end{eqnarray}
and, 
\begin{equation}
	\mathcal{V}_2 = \mu_{BS} Tr \left[ {\Phi^\dagger _B} \tilde \Phi_B\right]  \Phi_S^\ast + h.c. ,
\label{eq:pot_zero_tri-linear} 
\end{equation}
with \(\tilde{\Phi}_B \equiv \sigma_2 \Phi_B^{*} \sigma_2\). Throughout this work, all scalar potential parameters are taken to be real.  We further set \(v_2 = 0\), since this choice does not lead to phenomenologically distinct predictions compared to the case where both \(v_1\) and \(v_2\) acquire nonzero vacuum expectation values. In this limit, \(\mu_{BS} = 0\), as discussed in Ref.\,\cite{Bhattacharyya:2021lgr}. Therefore, our analysis of the phase transition is performed using the scalar potential \(\mathcal{V}_1\) alone. Furthermore, we take \(v_L = 0\), as a nonzero \(v_L\) would introduce an additional massless scalar degree of freedom that is phenomenologically undesirable.

The fermions of this model in terms of their chiral components can be written as:
\begin{equation}
\begin{gathered}
L_{L}=
\begin{pmatrix}
\nu_{L}\\
e_{L}
\end{pmatrix},
\qquad
L_{R}=
\begin{pmatrix}
\nu_{R}\\
e_{R}
\end{pmatrix},
\qquad
Q_{L}=
\begin{pmatrix}
u_{L}\\
d_{L}
\end{pmatrix},
\qquad
Q_{R}=
\begin{pmatrix}
u_{R}\\
d_{R}
\end{pmatrix}, \\[3mm]
Q_{LS}=q_{SL},
\qquad
Q_{RS}=q_{SR},
\qquad
L_{S}=l_{S},
\qquad
L_{B}=
\begin{pmatrix}
N_{1} & E_{1}\\
E_{2} & N_{2}
\end{pmatrix}.
\end{gathered}
\label{eq:fermions_chiral}
\end{equation}
Here, $L_L(L_R), Q_L(Q_R)$ are left handed (right handed)  lepton and quark doublets consisting of SM fermions along with right-handed neutrinos ($\nu_R$). $Q_{LS}$ and $Q_{RS}$ are color triplets and $SU(2)$ gauge singlet exotic quarks having $U(1)_L$ and $U(1)_R$ hyper-charges respectively.  They together form a $4$-component Dirac spinor $q_S$. $l_S$ is a neutral exotic singlet fermion carrying $U(1)_L$ and $U(1)_R$ gauge quantum numbers.  $N_1$ and $N_2$ are neutral heavy leptons while $E_1$ and $E_2$ are singly charged heavy leptons. They pair-wise form 4-component Dirac spinors, $N$ and $E$ respectively. 

The masses of the fermions arise from their Yukawa interactions with the Higgs fields. The relevant Yukawa Lagrangian is given by	
\begin{eqnarray}
	\mathcal{L}_{Y} &=& y_{qij} \bar{Q}_{iL} \Phi_B Q_{jR} + \tilde{y}_{qij} \bar{Q}_{iR} \tilde{\Phi}_B Q_{jL} + y_{lij} \bar{L}_{iL} \Phi_B L_{jR} + \tilde{y}_{lij} \bar{L}_{iR} \tilde{\Phi}_B L_{jL} + y_{sij} \bar{Q}_{iLS} \Phi_S Q_{jRS} \nonumber \\  &+& y_{LBij} \; Tr \left[ \bar{L}_{iB} \tilde{L}_{jB} \right] \Phi_S^c + \hat{y}_{LSij}\bar{L}_{iS} L_{jS}^c \Phi_S \Phi_S + y_{BBij}\; Tr\left[ \bar L_{iB} \tilde \Phi_B \right] L_{jS}^c + H.C.
\label{eq:lag_yukawa}
\end{eqnarray}
where, $i,j=1,2,3$ are generation numbers and $y$(s) are Yukawa coupling constants. $\Phi_S^c$ is complex conjugate of $\Phi_S$ and $\tilde{L}_B=\sigma_{2} L_B^* \sigma_{2}$. It is to be noted that among all the aforementioned Yukawa couplings, $\hat{y}_{LSij}$ is dimension full with canonical mass dimension $-1$. The corresponding dimension-$5$ mass term could have been generated via the scattering of a pair of singlet fermions 
on a pair of singlet higgs mediated by a heavy gauge boson or a scalar \cite{Bhattacharyya:2021lgr}. For simplicity, we have assumed the Yukawa matrices of all BSM fermions to be diagonal.
\section{Theoretical and Phenomenological constraints}
\label{sec:constraints}
For our subsequent analysis of the phase transition, the potential parameters appearing in Eq.~\eqref{eq:pot_zero} are required to satisfy a number of theoretical and phenomenological constraints. These theoretical and phenomenological bounds are summarized below.
\subsection{Vacuum Structure} \label{subsec:vacum_structure}
At the minimum of the scalar potential, the vacuum expectation values (VEVs) must satisfy the tree-level tadpole conditions,
\begin{equation}
\left.\frac{\partial V}{\partial \phi_i}\right|_{\rm vev}=0,
\qquad
\phi_i=\{\phi_1^0,\phi_2^0,\phi_L^0,\phi_R^0,\phi_S^0\}.
\end{equation}
As discussed in Section~\ref{sec:model}, we consider the vacuum alignment $v_2=0$ and $v_L=0$. Consequently, the five tadpole equations reduce to the following three independent relations:
\begin{eqnarray}
\mu_1^2 &=& \frac{1}{2}\left[2\lambda_1v_1^2+(c_1+c_3)v_R^2+\beta_1v_S^2\right], \nonumber\\
\mu_3^2 &=& \frac{1}{2}\left[(c_1+c_3)v_1^2+2\rho_1v_R^2+\gamma_1v_S^2\right], \nonumber\\
\mu_4^2 &=& \frac{1}{2}\left[\beta_1v_1^2+\gamma_1v_R^2+2\alpha_1v_S^2\right].
\label{eq:tadpole}
\end{eqnarray}
Taking $v_1$, $v_R$, and $v_S$ as input parameters, the above equations can be used to determine the dimensionful parameters $\mu_1^2$, $\mu_3^2$, and $\mu_4^2$. As a result, the scalar potential is described by ten independent dimensionless parameters: 
\begin{equation*}
\lambda_1,\ \lambda_3,\ \rho_1,\ \rho_3,\ c_1,\ c_3,\ c_4,\ \alpha_1,\ \beta_1,\ \gamma_1    
\end{equation*}
In the following, instead of  $c_1$, $c_3$ and  $c_4$,  very often the combination  $(c_1 + c_3)$ or $(c_1 + c_4)$ will arise in physical quantities. They will be denoted by $c_{13}$ and $c_{14}$ respectively.

The physical scalar spectrum must contain a Higgs boson with a mass of approximately $m_h \simeq 125~\mathrm{GeV}$ and couplings consistent with those of the Standard Model Higgs boson. Furthermore, the VEV of the bi-doublet is required to satisfy
\begin{equation}
\sqrt{v_1^2+v_2^2}=v_1\simeq246~\mathrm{GeV},
\end{equation}
where we have used the condition $v_2=0$. 
\subsection{Conditions for vacuum stability}\label{subsec:vacuum_stability}
To ensure the stability of the zero-temperature scalar potential in Eq.~\eqref{eq:pot_zero}, the potential must be bounded from below (BFB) in all field directions. For a scalar potential as complicated as ours, there is no straightforward analytical procedure that can completely guarantee the BFB conditions. Nevertheless, the copositivity criteria derived in~\cite{Kannike:2012pe} provide the necessary conditions for the scalar potential to satisfy vacuum stability. Following the method presented in~\cite{Chakrabortty:2013mha}, the co-positivity conditions for our model are given by
\begin{align}
\lambda_1 &\ge 0, \quad
(c_1 + c_3) + 2\sqrt{\lambda_1 \rho_1} \ge 0, \quad
(c_1 + c_4) + 2\sqrt{\lambda_1 \rho_1} \ge 0, \quad
\lambda_1 + \lambda_3 \ge 0, \quad
\rho_1 \ge 0, \nonumber \\
2\rho_1 &+ \rho_3 \ge 0, \quad
\alpha_1 \ge 0, \quad
\gamma_1 + 2\sqrt{\alpha_1 \rho_1} \ge 0, \quad
\beta_1 + 2\sqrt{\alpha_1 \lambda_1} \ge 0 ~.
\label{eq:bfb}
\end{align}
As discussed above, the copositivity criteria are necessary but not sufficient to guarantee that the scalar potential is bounded from below. Therefore, in addition to imposing these copositivity conditions, we numerically verified that all the parameter points used in this work satisfy the BFB conditions.
\subsection{Perturbative Unitarity}\label{subsec:unitarity}
The potential parameters should also obey the  perturbative unitarity constraints in longitudinal gauge boson and Higgs boson scattering processes. In the high energy limit, the most dominant contribution comes from the $s$-wave amplitude ($a_0$) at tree-level. The $S$-matrix unitarity for the scattering processes requires $|a_0| \leq 1$ which, implies that the eigenvalues  of the scattering submatrices: $ \leq 8 \pi$. Following the method given in~\cite{Mondal:2015fja,Arhrib:2011uy}, the unitarity conditions for this model setup are given by
\begin{align}
&4\lambda_1 \leq 8\pi,\quad
2(\lambda_1 + 2\lambda_3) \leq 8\pi,\quad
\lambda_1 - 2\lambda_3 + \rho_1 
\pm \sqrt{2(c_3 - c_4)^2 + (\lambda_1 - 2\lambda_3- \rho_1)^2}
\leq 8\pi, \nonumber \\
&c_1 + c_3 \leq 8\pi,\quad
c_1 + c_4 \leq 8\pi,\quad
c_1 + 2c_3 - c_4 \leq 8\pi,\quad
c_1 - c_3 + 2c_4 \leq 8\pi,\quad  4\rho_1 \leq 8\pi, \nonumber \\ 
&\rho_3 \leq 8\pi,\quad
\beta_1 \leq 8\pi,\quad
\gamma_1 \leq 8\pi,\quad 4\alpha_1 \leq 8\pi.
\label{eq:unitarity}
\end{align}
Some eigenvalues arsing from the scattering of neutral scalars do not have closed expressions. We have calculated those eigenvalues numerically and demanded the eigen values to be $\leq 8 \pi$. 
\subsection{Phenomenological Bounds}
\label{subsec:pheno_bounds}
One of the main phenomenological constraints on the model arises from the requirement that the physical scalar spectrum contains a neutral scalar with properties consistent with those of the observed SM Higgs boson, with a mass of approximately $m_h \simeq 125~\mathrm{GeV}$.

In general, the SM-like Higgs boson ($h^0$) is a linear combination of the neutral scalar fields $\{h_1^0,h_R^0,h_S^0\}$. 
The expressions for the masses of the physical scalars (after the electroweak symmetry breaking) are listed in Appendix A. 
To simplify the analysis, we impose the condition $ c_{13}=0$, which eliminates the mixing between the SM-like Higgs and the right-handed neutral Higgs scalar. We further set $\beta_1=0$, thereby removing the mixing between the singlet scalar and the SM-like Higgs.\footnote{As discussed in~\cite{Bhattacharyya:2021lgr}, the region with $\beta_1 > 0.001$ is already excluded by Higgs signal strength measurements.} In this limit, the SM-like Higgs mass is given by
\begin{equation}
    m_h^2 = 2\lambda_1 v_1^2,
\end{equation}
which fixes the quartic coupling $\lambda_1$ equals to $0.13$ for $v_1 = 246~\mathrm{GeV}$.

The lower bounds on the VEVs, \(v_R\) and \(v_S\), can be derived from the experimental constraints on the masses of the heavy charged and neutral gauge bosons, \(W_R\) and \(A'\), predicted by this model. The current lower limit on the \(W_R\) mass reported by the ATLAS Collaboration~\cite{ATLAS:2019isd} translates into the constraint \(v_R > 14.7~\mathrm{TeV}\). Likewise, searches for a heavy neutral gauge boson at the LHC~\cite{ATLAS:2017fih} place a lower bound on the mass of the \(A'\) boson, thereby constraining \(v_S\) for a given value of \(v_R\). For the minimum allowed value of \(v_R\), this constraint implies \(v_S > 12.61~\mathrm{TeV}\)~\cite{Bhattacharyya:2021lgr}. For the expresseions for the masses of the physical gauge bosons readers are referred to the Appendix A.

Coming back to the other scalars of the model, the lower bounds on charged scalars can be translated to combinations of some quartic parameters. As discussed in Ref.\cite{Bhattacharyya:2021lgr}, the charged scalar arising from the Left-handed doublet ($H^{\pm}_L$) has a lower bound of mass 494 GeV while the other charged scalar coming from bi-doublet ($H^{\pm}_1$) has to have a mass greater than 720 GeV to be consistent with LHC data. These in turn constrain the relevant quartic combinations as
\begin{eqnarray}
     (c_{14}-c_{13}) &> 0.005  \nonumber \\
     (\rho_3 - 2 \rho_1)& > 0.002.
\label{eq1:charged_higgs}
\end{eqnarray}
Similarly a lower bound of 800 GeV on the neutral scalar mass can loosely constrain the $\lambda_3$ parameter depending on the previous quartic relations. 

Before we close this section, let us mention that a Dirac fermion $N$ ($N_1$ and $N_2$ arising for the bi-doublet lepton $L_B$) and a Majorana fermion $L_s$ (see eq. 2.1) could be candidates for relic particles in the aforementioned BSM scenario. Their contribution to relic density and direct detection rate in experiments have been discussed in \cite{Bhattacharyya:2022trp}. It would be interesting to review the inelastic scattering  rate of $N \to L_s$ in the context of LZ experiment in light of their recently published result \cite{LZ:2026axp}.  Such a result will mainly constrain the mass difference between $N$ and $L_s$ along with the masses of heavy neutral gauge bosons ($A'$ and $Z'$) which mediates the interaction responsible for the scattering of DM over nucleons.  We restrain from doing such an exercise in the following  but will use the values of the VEVs which are consistent with the available LHC data.    
\section{Finite Temperature Effective Potential} \label{sec:effective_potential}
In order to study the phase transition in detail, we need to compute the finite-temperature effective potential. Since current experimental constraints require the LR symmetry breaking scale, \(SU(2)_{R} \otimes U(1)_{L} \otimes U(1)_{R} \rightarrow U(1)_{Y},\) to be much higher than the electroweak symmetry breaking scale, it is sufficient, to an excellent approximation, to consider only the \(\Phi_R\) and \(\Phi_S\) dependent effective potential when investigating the left-right phase transition.

The LR-breaking scale being much higher than the electroweak breaking scale, the LR phase transition takes place at a temperatures much higher than the electroweak scale. At such a temperature, the fields like $\phi_B$ and $\phi_L$ 
responsible for EWSB acquire positive thermal masses. Thus one may safely assume the background values of such fields to be practically vanishing during the LR phase transition. The relevant tunneling dynamics (between the false and true vacuum) can therefore be described in the reduced two-dimensional field space spanned only by $\phi^0_R$ and $\phi^0_S$,  which are responsible for LR symmetry breaking. In other words, for the purpose of studying the LR phase transition, the full finite-temperature effective potential can be evaluated in the subspace $ V_{\rm eff}(\phi_B,\phi_L,\phi_R,\phi_S;T) \;\rightarrow\; V_{\rm eff}(0,0,\phi_R,\phi_S;T) \equiv V_{\rm eff}(\phi_R,\phi_S;T)$. This reduction concerns only the background-field dependence of the effective potential. The degrees of freedom associated with $\Phi_B$ and $\Phi_L$  still contribute to the effective potential through their field-dependent masses and the corresponding zero- and finite-temperature loop corrections.

In general, the one-loop effective potential can be written as the sum of the tree-level potential, the Coleman--Weinberg (CW) correction and  the finite-temperature correction  with Daisy resummation~\cite{Quiros:1999jp}. However, quantum corrections shift both the position of the vacuum and the scalar masses from their tree-level values even at zero temperature. To preserve the tree-level vacuum expectation values and scalar masses at zero temperature and to cancel the zero temperature ultraviolet divergences, we include a counter term potential, $V_{\rm CT}$ following the prescription of Ref.~\cite{Graf:2021xku}. The full effective potential used in our analysis is therefore given by
\begin{equation}
    V_{\rm eff}(\phi_k, \mu, T) =
    V_{\rm tree}(\phi_k)+ V_{\rm CW}(\phi_k, \mu)  + \Delta V_T(\phi_k, T) + V_{\rm Daisy}(\phi_k, T) + V_{\rm CT}(\phi_k) ~. 
\label{eq:pot_eff}
\end{equation}
Here, \(\phi_k\) denote the background scalar fields. In our case, \(\phi_k={\phi_R,\phi_S}.\) The corresponding contributions to the effective potential are given by
\begin{equation}
    V_{\rm tree}(\phi_R,\phi_S) = - \frac{1}{2} \mu_3^2 \phi_R^2 - \frac{1}{2} \mu_4^2 \phi_S^2 + \frac{1}{4} \rho_1 \phi_R^4 + \frac{1}{4} \alpha_1 \phi_S^4 + \frac{1}{4} \gamma_1 \phi_R^2 \phi_S^2
\end{equation}
The one-loop CW contribution is given by
\begin{equation}
    V_{\rm CW}(\phi_R,\phi_S, \mu) = \sum_i \frac{n_i}{64\pi^2}\, m_i^4(\phi) \left[ \ln\!\left( \frac{m_i^2(\phi_R,\phi_S)}{\mu^2} \right) - c_i \right],
\label{eq:pot_cw}
\end{equation}
where the sum extends over all relevant scalar, gauge boson, and fermion degrees of freedom. Here, \(m_i^2(\phi_R,\phi_S)\) are the field-dependent squared masses given in Appendix~\ref{app:field_dependent_masses}, while \(n_i\) denotes the effective number of degrees of freedom of the \(i\)-th particle, taking into account its spin, color, and other internal degrees of freedom, In the \(\overline{\rm MS}\) renormalization scheme and Landau gauge, \(c_i=3/2\) for scalars and fermions, and \(c_i=5/6\) for gauge bosons.

To preserve the tree-level VEVs and scalar masses after inclusion of  the CW contribution, we introduce a set of counter terms as the following:
\begin{equation}
    V_{\rm CT}(\phi_R,\phi_S) = - \frac{1}{2} \delta \mu_3^2 \phi_R^2 - \frac{1}{2} \delta \mu_4^2 \phi_S^2 + \frac{1}{4} \delta \rho_1 \phi_R^4 + \frac{1}{4} \delta \alpha_1 \phi_S^4
    + \frac{1}{4} \delta \gamma_1 \phi_R^2 \phi_S^2 ~.
    \label{eq:V_CT}
\end{equation}
The coefficients of the counter term potential are determined by imposing the following renormalization conditions which are self evident:
\begin{equation}
\left.
\frac{\partial V_1}{\partial \phi_i}
\right|_{\phi=v_i}
= \left.
\frac{\partial^2 V_1}{\partial \phi_i \partial \phi_j}
\right|_{(\phi_i,\phi_j)= (v_i,v_j)} = 0,
\qquad (i,j=R,S),
\label{eq:renormalization_condition}
\end{equation}
where $V_1 = V_{\rm CW} + V_{\rm CT}$, and all the derivatives are calculated at the VEVs ($\phi_R = v_R, \phi_S = v_S)$. The explicit expressions for the CT coefficients are listed in the Appendix~\ref{app:field_dependent_masses}.

The determination of the counter terms from the vacuum renormalization conditions is affected by the well-known infrared divergence arising from massless Goldstone bosons. To regularize this spurious divergence, we introduce a small infrared regulator~\cite{Enomoto:2021dkl} in the Goldstone boson mass appearing in the CW potential. We have explicitly verified that the physical results are insensitive to the choice of the regulator.
The one-loop finite-temperature correction is given by
\begin{equation}
    \Delta V_T(\phi_R, \phi_S, T) = \frac{T^4}{2\pi^2} \sum_{i} n_i \, J_{i} \left(\frac{m_i^2(\phi_R,\phi_S)}{T^2}\right),
\label{eq:pot_thermal}
\end{equation}
where the sum extends over all particle species. The bosonic and fermionic thermal functions, \(J_B\) and \(J_F\), respectively, are defined as
\begin{align}
    J_{{B,F}}(a^2) &= \int_0^\infty dx\, x^2 \ln\!\left(1 \mp e^{-\sqrt{x^2+a^2}}\right).
\label{eq:thermal_int}
\end{align}
We also include the daisy (ring) resummation to regulate the infrared divergences associated with the bosonic zero Matsubara modes. To account for the leading infrared-sensitive contributions, we employ the Arnold--Espinosa prescription~\cite{Arnold:1992rz}, in which thermal masses are included only in the infrared-sensitive bosonic cubic term~\footnote{An alternative approach is the Parwani prescription~\cite{Parwani:1991gq}, in which the thermal masses are inserted into the full one-loop effective potential rather than only the infrared-sensitive cubic term.}
\begin{equation}
    V_{\rm Daisy}(\phi_R, \phi_S, T) = -\frac{T}{12\pi} \sum_{i} n_i \left[ \left(m_i^2(\phi)+\Pi_i(T)\right)^{3/2} - m_i^{3}(\phi) \right],
\label{eq:pot_daisy}
\end{equation}
where the sum runs over all scalar fields and the longitudinal components of the gauge bosons, which receive thermal (Debye) mass corrections. Here, \( \Pi_i(T) \) are the corresponding thermal self-energies. The relevant thermal masses calculated in our model are listed in the following:
\begin{align}
\Pi_{\Phi_B}(T)
&= T^2 \Bigg[\frac{g_{1L}^2+g_{1R}^2+27(g_{2L}^2+g_{2R}^2)}{144}
+ \frac{2y_{BB}^2+3y_t^2}{12}
+ \frac{10\lambda_1+4\lambda_3+\beta_1+2c_{13}+2c_{14}}{12}
\Bigg], \nonumber\\[2mm]
\Pi_{\Phi_L}(T)
&= T^2 \Bigg[
\frac{g_{1L}^2+4g_{1R}^2+27g_{2L}^2}{144}
+\frac{\gamma_1+6\rho_1+2\rho_3+2c_{13}+2c_{14}}{12}
\Bigg], \nonumber\\[2mm]
\Pi_{\Phi_R}(T)
&= T^2 \Bigg[
\frac{4g_{1L}^2+g_{1R}^2+27g_{2R}^2}{144}
+\frac{\gamma_1+6\rho_1+2\rho_3+2c_{13}+2c_{14}}{12}
\Bigg], \nonumber\\[2mm]
\Pi_{\Phi_S}(T)
&= T^2 \Bigg[
\frac{g_{1L}^2+g_{1R}^2}{36}
+\frac{2y_{LB}^2+3y_s^2}{12}
+\frac{\alpha_1+\beta_1+\gamma_1}{3}
\Bigg], \nonumber\\[2mm]
\Pi_{W_R}(T)
&=\frac{7}{3}g_{2R}^2T^2, \qquad
\Pi_{B_L}(T)=\frac{13}{27}g_{1L}^2T^2,\qquad
\Pi_{B_R}(T)=\frac{13}{27}g_{1R}^2T^2.
\label{eq:themal_mass}
\end{align}
\section{Left-Right Phase Transition  and Symmetry Breaking Pattern} \label{sec:phase_transition}
We are now equipped with necessary ingredients to study the phase transition. The temperature-dependent VEVs, $v_i(T)$, can be realized by minimizing the effective potential as given in Eq.~\eqref{eq:pot_eff}. The nature of the phase transition will be determined by the existence of two degenerate minima, separated by a potential barrier at a particular temperature known as the critical temperature, $T_c$. If such degenerate minima exist, the phase transition is first order. Otherwise, the transition is either second order or a crossover.

In the two-dimensional field space spanned by the scalar fields $\phi_R {~\rm and} ~\phi_S$, the thermal evolution of the vacuum can proceed via several ways as the Universe cools. Starting from the fully symmetric phase at $(v_R = 0,v_S =0)$ and eventually reaching the fully broken  phase at $(v_R,v_S)$, the system can undergo two qualitatively distinct types of phase transitions: a two-step phase transition or an one-step phase transition. In the two-step scenario, the Universe first transitions to an intermediate vacuum, followed by a second transition to the fully broken vacuum $(v_R,v_S)$. Thus, the symmetry breaking occurs sequentially, with the two scalar fields acquiring their vacuum expectation values at different temperatures. In contrast, in the one-step scenario, both fields acquire nonzero vacuum expectation values simultaneously, and the system transitions directly from $(0,0)$ to $(v_R,v_S)$.

\begin{figure}[!htbp]
  \centering
  \includegraphics[width=0.7\linewidth]{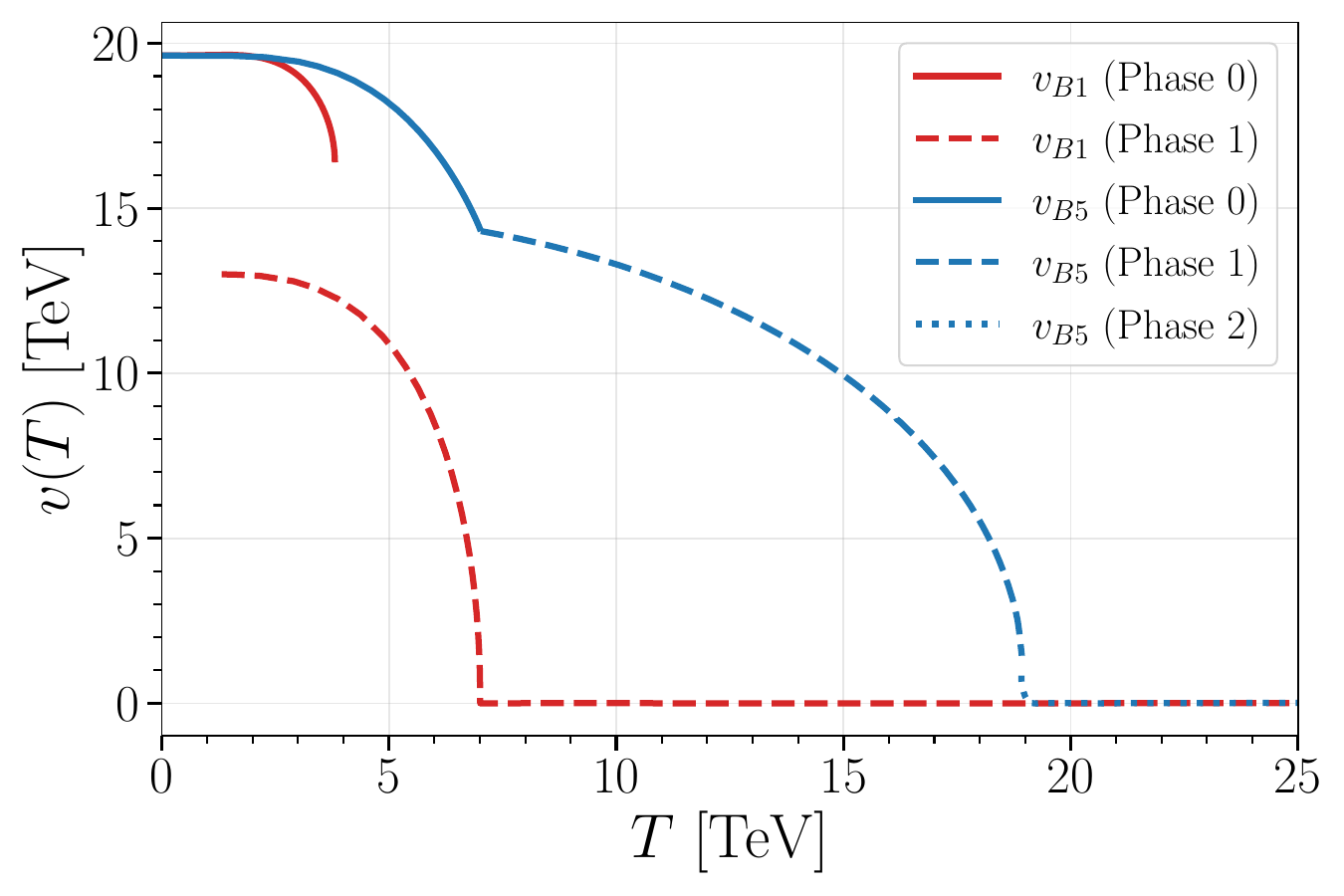}
  \caption{Temperature dependence of the order parameter $v(T)=\sqrt{v_R(T)^2+v_S(T)^2}$ for two benchmark points: BP-1 (red) and BP-5 (blue).}
  \label{fig:temp_order}
\end{figure}

The actual phase-transition pattern is determined by the temperature dependence of the effective potential~\eqref{eq:pot_eff}, which determines the location and relative depth of the local minima. At high temperatures, thermal corrections generally stabilize the origin, making $(0,0)$ the global minimum. As the temperature decreases, additional minima may develop. We will try to illustrate this in Fig.~\ref{fig:temp_order} which  shows the thermal evolution of $v(T) = \sqrt{v_R ^2 (T) + v_S ^2 (T)}$ for two sets of  model parameters. At high temperatures, both the field approaches zero, corresponding to the symmetric phase. As the Universe cools, the filed acquires a non-zero value, signaling the transition to a symmetry-broken phase. Two different sets (BP-1 and BP-5, for the exact values of of the parameters see Table.~\ref{tab:bp}) of model parameters are used show the pattern of symmetry breaking. 

The BP-1 benchmark exhibits two phase branches clearly indicating a first order transition near $T = 3.5 ~\rm TeV$. Three distinct branches are present for BP-5. Transition around $T = 6 ~\rm TeV$ corresponds to a cross-over of phase transition which is second order in nature. These branches trace the temperature evolution of the corresponding minima of the effective potential and illustrate the different thermal histories of the two benchmark scenarios. The dissimilarities in the temperature evolution of order parameter ($v$) for two benchmark points could be mainly attributed to the parameter $\gamma_1$ which is responsible for the coupling between two scalar fields $\phi_R^0$ and $\phi_S^0$. For BP-1, $\gamma_1$ being zero, $v_R$ and $v_S$ evolve independently of each other, with $T$. This feature will be illuminated in the next section in some details. 

Owing to the multidimensional scalar potential of this model, all of these types of phase transitions can occur for different choices of the scalar potential parameters. Since our primary interest is the generation of observable GW signals from LR symmetry breaking, we focus on the region of parameter space where the phase transition is first order.

A first-order phase transition proceeds through the nucleation of bubbles of the true vacuum, characterized by $v_i(T) \neq 0$, within the metastable false vacuum. These bubbles subsequently expand and convert the surrounding false-vacuum phase into the true-vacuum phase~\cite{Linde:1977mm,Linde:1981zj}.    
The scalar field can transit from the false vacuum to the true vacuum either via quantum tunneling or via thermal fluctuation. In general, the tunneling rate per unit volume can be expressed as~\cite{Coleman:1977py}
\begin{equation}
    \Gamma(T)=A(T)\exp[-S(T)]\,,
    \label{eq:nucleation_rate}
\end{equation}
where $A(T)$ is the fluctuation prefactor with mass dimension four, and $S(T)$ is the Euclidean bounce action. The form of the bounce action depends on the temperature. At zero temperature, quantum tunneling is dominated by the $O(4)$-symmetric bounce solution, and the tunneling rate is given by
\begin{equation}
    \Gamma_4 \simeq A_4 e^{-S_4},
\end{equation}
where $S_4$ is the four-dimensional Euclidean action evaluated on the bounce solution.
At finite temperature, provided the temperature is sufficiently high compared to the inverse bubble size, the Euclidean time direction (or equivalently $1/T$ direction) becomes compact, and the dominant tunneling configuration becomes $O(3)$ symmetric with $ S\equiv S_3(T)/T $. In this case, the tunneling rate takes the form
\begin{equation}
    \Gamma_3(T)\simeq A(T)\exp\!\left[-\frac{S_3(T)}{T}\right],
    \label{def-gamma3}
\end{equation}
where $S_3(T)$ is the three-dimensional Euclidean action of the critical bubble, given by:
\begin{equation}
    S_3(T) = 4\pi \int_0^\infty \dd r \: r^2 \left[\frac{1}{2}\left(\frac{\dd\phi_k}{\dd r}\right)^2+\Delta V_{\rm eff}(\phi_k,T)\right] \, ,
\label{eq:s3}
\end{equation}
where $\Delta V_{\rm eff}(\phi_k,T) \equiv V_{\rm eff}(\phi_k,T) - V_{\rm eff}(\phi_{k \rm false},T)$. The $\phi_k$ corresponds to the solutions of equation of motion, 
\begin{equation}
    \frac{\dd^2 \phi_k}{\dd r^2} + \frac{2}{r} \frac{\dd\phi_k}{\dd r} - \frac{\partial V_{\rm eff}}{\partial \phi_k}= 0 \, ,
\label{eq:eom}
\end{equation} 
with the boundary conditions
\begin{equation}
    \frac{\dd\phi_k}{\dd r}(r=0)=0 \, , \qquad \phi_k(r=\infty)=0 \, .
\label{eq:boundary}
\end{equation}
At finite temperature,  the bubble nucleation rate, $\Gamma_3$ can be expressed as:
\begin{equation}
    \Gamma_3 \simeq T^4\left(\frac{S_3(T)}{2\pi T}\right)^{3/2} \exp\left[-\frac{S_3(T)}{T}\right] \,.
\label{eq:gamma_3}
\end{equation}
There are several numerical methods and publicly available packages for solving the bounce equation~\eqref{eq:eom} and subsequently evaluating the Euclidean action~\eqref{eq:s3}. In the single-field case, the overshoot/undershoot shooting method~\cite{Coleman:1977py} is widely used due to its simplicity and robustness. However, for theories involving multiple scalar fields, the problem becomes significantly more challenging because the tunneling trajectory in field space is not known a priori. Various algorithms have therefore been developed to determine the bounce solution in multi-field potentials, including path deformation methods, string methods, and minimization techniques. Several publicly available packages implement these algorithms, such as \texttt{CosmoTransitions}~\cite{Wainwright:2011kj}, which employs the path deformation algorithm, \texttt{AnyBubble}~\cite{Masoumi:2016wot}, which is based on the multiple-shooting method, \texttt{BubbleProfiler}~\cite{Athron:2019nbd}, which implements perturbative and direct shooting techniques, and \texttt{FindBounce}~\cite{Guada:2020xnz}, which uses a gradient-flow approach. In this work, we employ the \texttt{CosmoTransitions} package to determine the bounce solutions and evaluate the three-dimensional Euclidean action.

As the phase transition and bubble nucleation occur during the very early stages of the Universe, the cosmic expansion rate plays an important role and must therefore be taken into account. During the radiation-dominated era, the expansion rate is characterized by the Hubble parameter,
\begin{equation}
    H(T)\simeq 1.66\sqrt{g_*}\,\frac{T^2}{M_{\rm Pl}} \, ,
\end{equation}
where $g_*\simeq 155$ denotes the effective number of relativistic degrees of freedom for this model, and $M_{\rm Pl}\simeq 2.4\times10^{18}\,\mathrm{GeV}$ is the reduced Planck mass.

For the phase transition to complete successfully, the bubble nucleation rate must be sufficiently large compared to the expansion rate of the Universe. If the expansion rate exceeds the bubble nucleation rate, bubbles of the true vacuum cannot nucleate efficiently, preventing the phase transition from completing. The nucleation temperature, $T_n$, is therefore defined as the temperature at which, on average, one critical bubble is nucleated within a Hubble volume during one Hubble time, {\it i.e.},
\begin{equation}
    \left.\frac{\Gamma(T)}{H^4(T)}\right|_{T=T_n}=1 \, ,
    \label{eq:tn}
\end{equation}
and the nucleation rate $\Gamma$ is defined in Eq.\eqref{def-gamma3}. A closely related characteristic temperature is the percolation temperature, $T_p$, defined as the temperature at which approximately one-third of the physical volume has transitioned to the true vacuum. For a precise definition, we refer the reader to Ref.~\cite{Ellis:2018mja}.
\section{Numerical Analysis}\label{sec:na}
We are now ready to report the results of a numerical scan of the parameter space to check whether a sufficiently SFOPT takes place in the framework of our interest. To this end, the following parameter values have been chosen after imposing all the theoretical and phenomenological constraints discussed in Section~\ref{sec:constraints}.
\begin{align}
c_{14} \in [0, 2],\quad
\rho_{1} \in [0, 0.5],\quad
\rho_{3} \in [1, 2], \quad
\lambda_{3} \in [0, 2],\quad
\gamma_{1} \in [0,2],\quad
\alpha_{1} \in [0,2] 
\label{eq:scan_range}
\end{align}
Fixing $v_R = 14.7~\mathrm{TeV}$ and $v_S = 13~\mathrm{TeV}$, we reproduce the observed properties of the SM Higgs boson by fixing $v_1 = 246~\mathrm{GeV}$, $\lambda_1 = 0.13$, and $\beta_1 = 0$. In the next section, we will briefly state numerical result of our analysis  with non-zero $\beta_1 > 0$. For simplicity, we also set $c_{13} = 0$, as discussed above. In choosing the parameter ranges, we take into account the relevant phenomenological constraints. In  particular, we choose the ranges of $\rho_1$ and $\rho_3$ such that the squared masses of the physical charged Higgs bosons remain positive, requiring $\rho_3 > 2\rho_1$ throughout the scan. The Yukawa matrices for the exotic fermions are diagonal with diagonal entries equal to unity. Consequently their masses are equal to the VEVs ($v_R$ or $v_s$) which generate them.

\begin{figure}[H]
    \centering
    \includegraphics[width=1.0\linewidth]{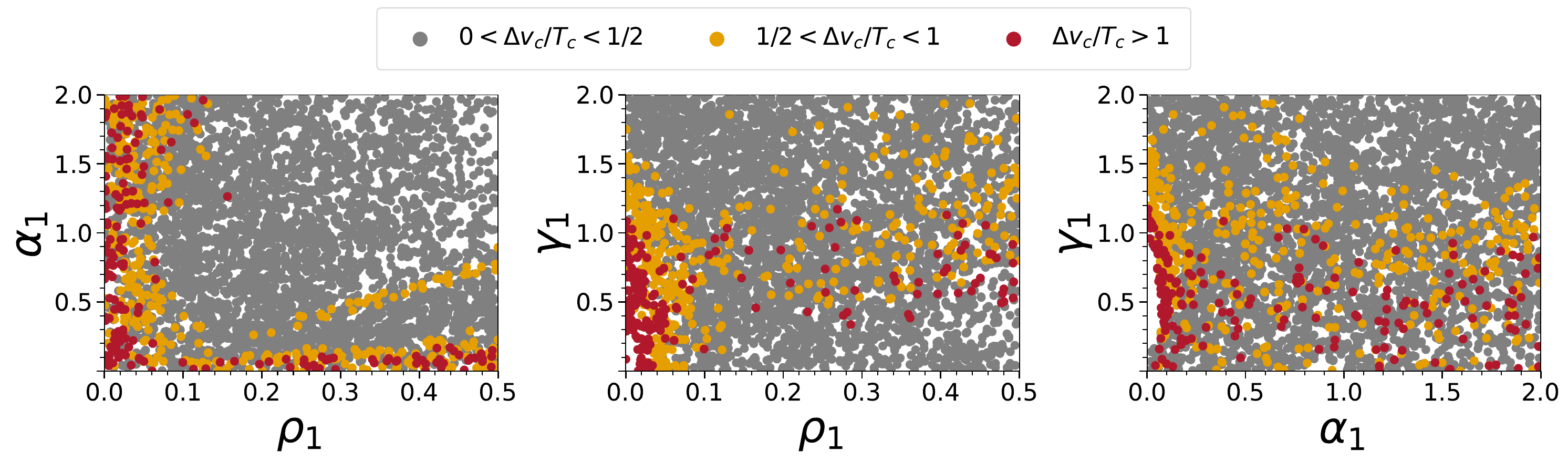}
    \caption{Strength of the LR-symmetry-breaking phase transition, characterized by $\Delta v_c/T_c$. The figure shows randomly generated two-dimensional projections of the full parameter space within the ranges specified in Eq.~\eqref{eq:scan_range}. The different colors indicate different ranges of the phase-transition strength parameter $\Delta v_c/T_c$, while the red dots denote points corresponding to a SFOPT.}
    \label{fig:scan_vctc}
\end{figure}

\begin{figure}[htbp]
    \centering
    \includegraphics[width=\linewidth]{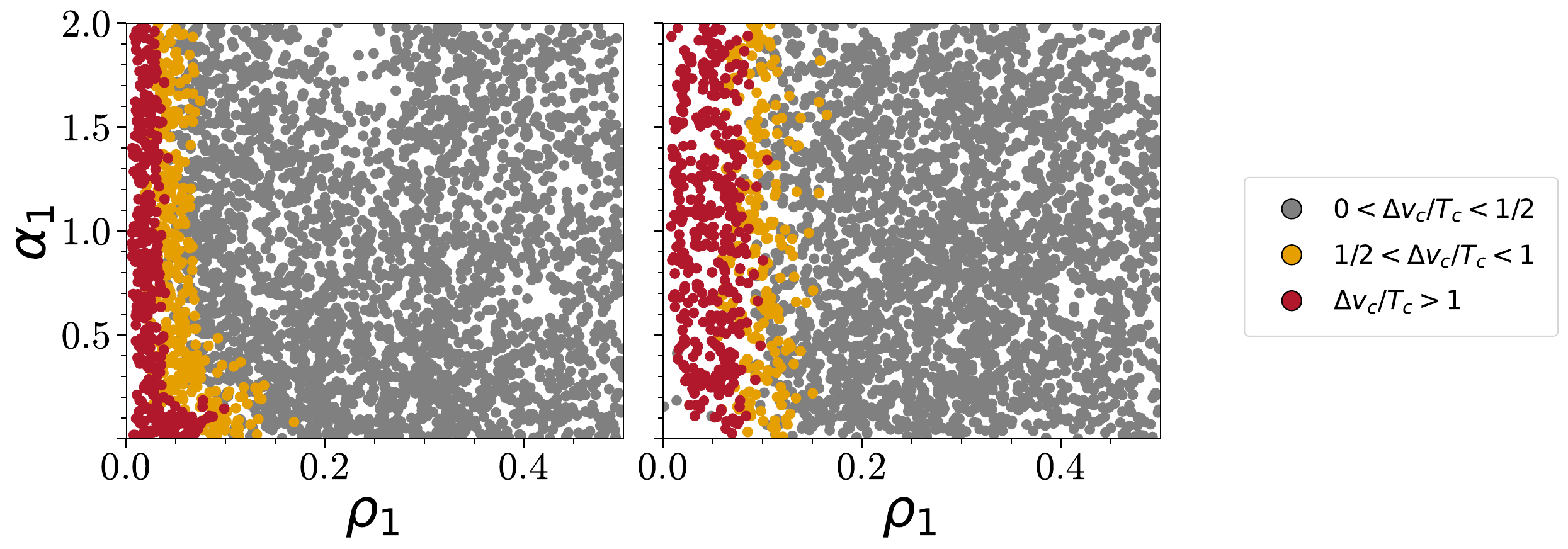}
    \caption[Strength of the LR-symmetry-breaking phase transition for fixed $\gamma_1$]{Strength of the LR-symmetry-breaking phase transition, characterized by $\Delta v_c/T_c$. In this scan, the mixing parameter between $\Phi_R$ and $\Phi_S$ is fixed. The left panel corresponds to $\gamma_1=0$, while the right panel corresponds to $\gamma_1=0.1$. The remaining parameters are randomly generated within the ranges specified in Eq.~\eqref{eq:scan_range}.} 
    \label{fig:scan_vctc_gamma}
\end{figure}

A large sample of random parameter points have been generated within the specified parameter ranges such that  the LR-symmetry-breaking phase transition is of first order for such values of the parameters. Since calculating the phase-transition strength parameter $\alpha$ for a large number of parameter points is computationally expensive, we use the ratio $\Delta v_c/T_c$ as a measure of the transition strength. $\Delta v_c$ denotes the difference of order parameter corresponding to the broken and unbroken phases at the critical temperature $T_c$. A commonly used criterion $\frac{\Delta v_c}{T_c} \geq 1$ has been imposed to identify a SFOPT. The results are presented in Figs.~\ref{fig:scan_vctc} and \ref{fig:scan_vctc_gamma}. Red dots on the aforementioned plots indicate the parameter points for which one can realize SFOPT. Such parameter points mostly correspond to relatively small values $(\leq 1)$ of the mixing parameter $\gamma_1$ between $\phi^0_R$ and $\phi^0_S$. On the other hand, it is evident from the scatter plots that occurrence of SFOPT is nearly independent of $\alpha_1$, the self quartic coupling of $\phi^0_S$. We will try to explain this briefly and qualitatively in the following. 
 
The preference for relatively small $\gamma_1$ can be qualitatively understood from the structure of the scalar potential. In the limit $\gamma_1\rightarrow0$, the $\phi^0_R$ and $\phi^0_S$ directions become approximately decoupled, allowing the phase transition to proceed predominantly along the direction with the more favorable thermal barrier. Increasing $\gamma_1$ couples the two directions through the $(\phi^0_R\phi^0_S )^2$ interaction and modifies the effective quartic coupling along a mixed tunneling trajectory, which can reduce the relative importance of the thermal barrier. Consequently, smaller values of $\gamma_1$ tend to favor a stronger FOPT by allowing the tunneling trajectory to remain closer to the direction with the more efficient thermal barrier. 
Let us now try to understand the role of $\rho_1$ and $\alpha_1$ in determining the finite-temperature effective potential. A scan in the parameter space is performed after fixing the values of the VEVs and the mixing parameter $\gamma_1$. It has been observed that that the phase transition is more likely to be strongly first order for values of the quartic coupling $\rho_1$ to be less than $0.1$. No such pronounced dependence on $\alpha_1$ is observed. Such a behavior can be qualitatively understood from the formation of the thermal barrier responsible for the first-order transition. The barrier receives an important contribution from the bosonic cubic terms in the finite-temperature effective potential, particularly from the gauge-boson sector. In our model, the gauge-boson masses have a stronger dependence on $\phi^0_R$ than on $\phi^0_S$, resulting in a larger gauge-induced cubic contribution along the $\phi^0 _R$ direction. Consequently, reducing $\rho_1$ enhances the relative importance of the thermal barrier and more readily strengthens the phase transition, whereas the corresponding effect of reducing $\alpha_1$ is less pronounced. When a nonzero $\gamma_1$ is introduced, the two scalar directions become coupled through the portal interaction $(\phi_R^0\phi_S^0)^2$, allowing the tunneling trajectory to acquire a mixed but still $\phi^0_R$-dominated character. As a result, the stronger gauge-induced barrier associated with the $\phi^0_R$ sector can remain effective even for relatively larger values of $\alpha_1$, explaining the observed shift of the SFOPT region towards larger $\alpha_1$ as $\gamma_1$ increases, as shown in Fig.~\ref{fig:scan_vctc_gamma}. This provides a qualitative explanation of the observed scan behavior, while the precise strength and dynamics of the phase transition are determined by the complete one-loop finite-temperature effective potential, including scalar-loop contributions, thermal masses, daisy re-summation, and the multidimensional tunneling dynamics.

We have chosen eight set of benchmark points (BP-1 - BP-8) for our following analysis of GWs. While choosing the benchmark points, the case of zero and non-zero values of the mixing $\gamma_1$, the hierarchy in the two quartic parameters ($\rho_1$ and $\alpha_1$), VEVs ($v_S$ and $v_R$) have been taken into consideration. The exact values of the parameters for different benchmark points are listed in Table \ref{tab:bp}. It is needless to be mentioned that for all such points $\frac{\Delta v_c}{T_c} \geq 1$ hallmarking the nature of a SOFPT. 
\begin{table}[H]
    \centering
    \begin{tabular}{|c|c|c|c|c|c|c|c|c|}
    \hline 
       BP & $v_R $ \small (TeV) & $v_S $ \small (TeV) & $c_{14}$ & $\alpha_1 $ & $ \rho_1 $  & $ \rho_3 $   & $\lambda_{3}$ & $\gamma_{1}$   \\
        \hline
        BP-1 & 14.70 & 13.00 & 0.40 & 0.10 & 0.01 & 1.00 & 0.01 & 0.00 \\
        \hline
        BP-2 & 14.70 & 13.00 & 0.55 &1.78 & 0.05 & 1.95 & 0.40 & 0.38 \\
        \hline
        BP-3 & 14.70 & 13.00 & 1.40 & 0.10 & 0.05 & 1.50 & 1.00 & 0.50 \\
        \hline
        BP-4 & 14.70 & 13.00 & 0.50 & 0.10 & 0.02 & 1.40 & 0.50 & 0.01 \\
        \hline
        BP-5 & 14.70 & 13.00 & 0.80 & 0.10 & 0.30 & 1.00 & 0.01 & 0.02 \\
        \hline
        BP-6 & 15.83 & 65.55 & 0.40 & 0.10 & 0.01 & 1.00 & 0.01 & 0.00 \\
        \hline        
        BP-7 & 49.70 & 29.74 & 0.40 & 0.10 & 0.01 & 1.00 & 0.01 & 0.00 \\
        \hline        
        BP-8 & 14.97 & 68.18 & 0.40 & 0.10 & 0.01 & 1.00 & 0.01 & 0.00 \\
      \hline
    \end{tabular}
    \caption{Benchmark values of the scalar potential parameters for our analysis. The value of SM vev $v$ is kept fixed at $246 ~\rm GeV$ and $\lambda_1 = 0.13$, $\beta_1 = 0$.}
    \label{tab:bp}
\end{table}
\section{Gravitational Wave Spectrum}\label{sec:GW}
A first-order phase transition generates SGWB and conventionally characterized by the GW energy density spectrum as
\begin{equation}
   h^2 \Omega_{\rm GW}(f) \equiv
    \frac{h^2}{\rho_c} \frac{\dd\rho_{\rm GW}}{\dd\ln f} \, ,
    \label{eq:gw_amplitude}
\end{equation}
where $h\simeq0.7$ is the dimensionless Hubble parameter, $\rho_{\rm GW}$ is the energy density stored in gravitational waves, and
    $\rho_c ~( \simeq
    1.05\times10^{-5}\,h^2~{\rm GeV}/{\rm cm}^3)$
is the critical energy density of the Universe. The factor of $h^2$ is conventionally included so that the quoted GW spectrum is independent of the experimental uncertainty in the present-day Hubble parameter $H_0$~\cite{DiValentino:2021izs}.

In a first-order phase transition, SGWB is generated by the expansion and collision of true-vacuum bubbles and their interactions with the surrounding thermal plasma. The GW signal receives contributions from three primary sources: collisions of expanding bubble walls ($\Omega_c$)~\cite{
Weir:2016tov,Jinno:2017fby,Jinno:2019bxw,Lewicki:2020jiv,Megevand:2021juo}, long-lasting sound waves in the bulk plasma ($\Omega_s$)~\cite{
Hindmarsh:2016lnk,Hindmarsh:2017gnf,Hindmarsh:2019phv}, and magneto-hydrodynamics (MHD) turbulence in the plasma ($\Omega_t$)~\cite{
Kisslinger:2015hua,RoperPol:2019wvy}. The total GW spectrum can therefore be approximated as the sum of these three contributions,
\begin{equation}
    h^2\Omega_{\rm GW}(f) \simeq h^2\Omega_c(f) +h^2\Omega_s(f) + h^2\Omega_t(f) \, .
    \label{eq:gw_total}
\end{equation}
These three contributions can be parameterized in a model-independent way in terms of a couple of dimensionless SFOPT parameters: the transition strength $\alpha$ and the inverse duration parameter $\beta/H_*$. The parameter $\alpha$ is defined as the ratio of the vacuum energy density $(\epsilon_*)$ released during the phase transition to the radiation energy density $(\rho_{\rm rad} =\frac{\pi^2}{30}\,g_*\,T_*^4)$,
\begin{equation}
    \alpha=\frac{\epsilon_*}{\rho_{\rm rad}} \, ,
     \label{eq:alpha}
\end{equation}
The parameter $\beta/H_{*}$ characterizes the inverse duration of the phase transition and is defined by
\begin{equation}
    \frac{\beta}{H_*} =
    \left. -\frac{T}{\Gamma} \frac{\dd\Gamma}{\dd T} \right|_{T=T_*} \, ,
    \label{eq:beta}
\end{equation}
All the quantities are evaluated at a reference temperature $T=T_*$, where $T_*$ denotes either the nucleation temperature $T_n$ or the percolation temperature $T_p$~\footnote{For the calculation of the GW spectrum, we take the percolation temperature $T_p$ as the reference temperature. The strength parameter $\alpha$ is evaluated at $T_p$, as this provides a more appropriate characterization of the transition at the temperature relevant for GW production. However, we find that the determination of the inverse duration parameter $\beta/H_*$ at the nucleation temperature $T_n$ is numerically more stable.}. 

Although the \texttt{CosmoTransitions} package has been used to solve the bounce equation in Eq.~\eqref{eq:s3}, we do not rely on its built-in approximation, $S_3(T_n)/T_n \sim 140$, to determine the nucleation temperature. Instead, we calculate the nucleation temperature using our own numerical implementation. Among the different phase-transition quantities, the determination of $\beta/H_*$ is particularly sensitive to the numerical evaluation of the bounce action. To obtain a reliable estimate, we calculate the action $S_3(T)$ in a small temperature interval around the transition temperature $T_*$ and perform a linear fit to the resulting values of $S_3(T)/T$. The slope obtained from this fit is then multiplied by $T_*$ to determine $\beta/H_*$, as given in Eq.~\eqref{eq:beta}. This procedure provides a numerical estimate of $\beta/H_*$ while reducing the sensitivity to numerical fluctuations in the individual values of the bounce action.

In Table \ref{tab:pt_par}, we present the calculated values of $\alpha$, $\frac{\beta}{H_*}$, $T_c$, $T_n$ and $T_p$ for bench marks values of the parameters. These quantities are necessary ingredients for calculating the GW spectrum.

We would like to comment about the effect of considering non-zero values of $\beta_1$. For the set of benchmarks, we found out that for non-zero allowed values of $\beta_1 (< 0.001)$, the changes in the critical and nucleation temperature are minuscule for all the BPs except BP-3. The reason is possibly the appearance of large non-zero value of $\gamma_1$, so first order phase transition vanishes when another mixing $\beta_1$ is turned on.
\begin{table}[H]
    \centering
    \begin{tabular}{|c|c|c|c|c|c|c|c|c|}
    \hline 
       BP  & $T_c$ \small [GeV]  
       & $T_n $ \small [GeV] & $T_p$ \small [GeV]  &  $\alpha $ & $ \beta/H_{*} $    \\
       \hline
       BP-1 & 3355.26  & 1564.74   &  1481.84  & 0.41129  &  269.29 \\
       \hline
       BP-2 & 5428.70  & 3010.57  & 2789.82 & 0.16188 & 239.28 \\
      \hline
      BP-3 & 19696.30  &  19555.68   & 19360.13  & 0.00715 &  3097.76 \\
      \hline
      BP-4 & 4535.72  & 2771.74   & 2631.23 & 0.14367  &  309.82\\
      \hline
      BP-5 & 18920.72 &  18920.35  &  18920.16  &  0.00006  &  175174.98 \\
     \hline
      BP-6 & 3661.20  & 1772.65   & 1691.09  & 0.35486  &  488.93 \\
     \hline
     BP-7 & 11336.26  & 5225.67   &  4984.47  & 0.41893  &  323.97 \\
    \hline
     BP-8 & 3361.93 & 1435.62   &  1367.78 & 0.53076  &  157.32 \\
    \hline
    \end{tabular}
    \caption{The values of phase transition parameters for sets of benchmark points.}
    \label{tab:pt_par}
\end{table}
Apart from $\alpha$ and $\beta/H_{*}$, the shape and the normalization of the spectrum also depend on  the transition temperature $T_*$, the bubble-wall velocity $v_w$, and the efficiency factors, $\kappa$ that determine the fraction of the released vacuum energy transferred to the scalar field and the surrounding plasma. To a very good approximation, the bubble-wall velocity in the plasma is taken to be 1 (in units of $c$) in the present analysis. The contribution from the acoustic waves which determines the peak frequency of the signal, can be parametrized as 
\begin{equation}
 h^2\Omega_s = 2.65\times 10^{-6} \; \Upsilon(\tau_{\rm s})\left(\frac{v_w}{\beta/H_*}\right)\left(\frac{100}{g_*(T_*)}\right)^{1/3}\left(\frac{\kappa_s \alpha}{1+\alpha}\right)^2 \left(\frac{f}{f_s}\right)^3\left[\frac{7}{4+3(f/f_s)^2}\right]^{7/2},  \\
\end{equation}
The factor, \(\Upsilon(\tau_{\rm s})\) accounts for the finite lifetime of the sound-wave period~\footnote{Numerical simulations of the acoustic phase have revealed that the source of the sound-wave is active only for a finite duration of time. This results in a suppression of the GW amplitude relative to the idealized case of a source lasting for Hubble time scale. Such an effect is accounted by the suppression factor \(\Upsilon(\tau_{\rm s})\)} and is defined by~\cite{Guo:2020grp}
\begin{equation}\label{eq:swtimepart}
\Upsilon(\tau_{\rm s}) = 1 - \frac{1}{\sqrt{1+2 \tau_{\rm s} H_{\ast}}},
\end{equation}
Following Ref.~\cite{Hindmarsh:2017gnf}, the sound-wave lifetime $\tau_{\rm s}$ is approximately equal to $R_N/\overline{U}_f$, where the mean bubble separation $R_* = (8\pi)^{1/3}v_w \beta_*^{-1}$ and the root-mean-squared fluid velocity $\overline{U}_f=\sqrt{3\kappa_{\rm s}\alpha/4}$. The sound efficiency factor $\kappa_s$ is given by~\cite{Kamionkowski:1993fg}
\begin{equation}
\label{eq:kappasw}
\kappa_{\rm s} =
\frac{\sqrt{\alpha}}
{0.135+\sqrt{0.98+\alpha}}.
\end{equation}
The peak frequency of sound wave can be written as,
\begin{equation}
  f_s = 1.9\times10^{-2}~\textrm{mHz} \left(\frac{g_*(T_*)}{100}\right)^{1/6}\left(\frac{T_*}{100~\GeV}\right)\left(\frac{\beta/H_*}{v_w}\right).      
\end{equation}
The contribution from the bubble collision term is given by~\cite{Lewicki:2021pgr}
\begin{equation}
 h^2\Omega_c = 1.67\times 10^{-5} \left(\frac{v_w}{\beta/H_*}\right)^2\left(\frac{100}{g_*(T_*)}\right)^{1/3}\left(\frac{\kappa_c \alpha}{1+\alpha}\right)^2\left(\frac{0.11v_w}{0.42 + v^2_w}\right)\left(\frac{f}{f_c}\right)^{2.8}\left[\frac{3.8}{1+2.8(f/f_c)^{3.8}}\right], 
 \label{eq:omega}
\end{equation}
where the factor $k_c$ stands for the fraction of released vacuum energy that is converted into scalar-field gradient energy and  is given by ~\cite{Kamionkowski:1993fg}
\begin{equation}
\label{eq:kcfac}
\kappa_c =
\frac{0.715\,\alpha + \dfrac{4}{27}\sqrt{\dfrac{3\alpha}{2}}}
{1 + 0.715\,\alpha}.
\end{equation}
and the peak frequency in this case, is given by
\begin{equation}
    f_c = 1.6\times10^{-2}~\textrm{mHz} \left(\frac{g_*(T_*)}{100}\right)^{1/6}\left(\frac{T_*}{100~\GeV}\right)\left(\frac{\beta/H_*}{v_w}\right)\left(\frac{0.62 v_w}{1.8 - 0.1 v_w + v^2_w}\right).
\end{equation}
Finally, the contribution from the magneto-hydrodynamic turbulence to the gravitational wave spectrum can be parametrized by 
\begin{equation}
  h^2\Omega_t = 3.35\times 10^{-4} \left(\frac{v_w}{\beta/H_*}\right)\left(\frac{100}{g_*(T_*)}\right)^{1/3}\left(\frac{\epsilon \;\kappa_s \alpha}{1+\alpha}\right)^{3/2} \left(\frac{f}{f_t}\right)^3\left[\frac{1}{1+(f/f_t)}\right]^{11/3}\left(\frac{1}{1+8\pi f/h_*}\right)\,
\end{equation}
where $\epsilon$ denotes the fraction of the bulk kinetic energy of the plasma that is converted into turbulence and the precise numerical value of $\epsilon$ is not well known. Numerical simulations indicate that $\epsilon \simeq 0.1$~\cite{Caprini:2015zlo,Hindmarsh:2015qta}, corresponding to $\kappa_{\rm t}\simeq 0.1\,\kappa_{\rm s}$ which is adopted throughout in our numerical analysis.

\begin{figure}[H]
\centering
\includegraphics[width=0.7\linewidth]{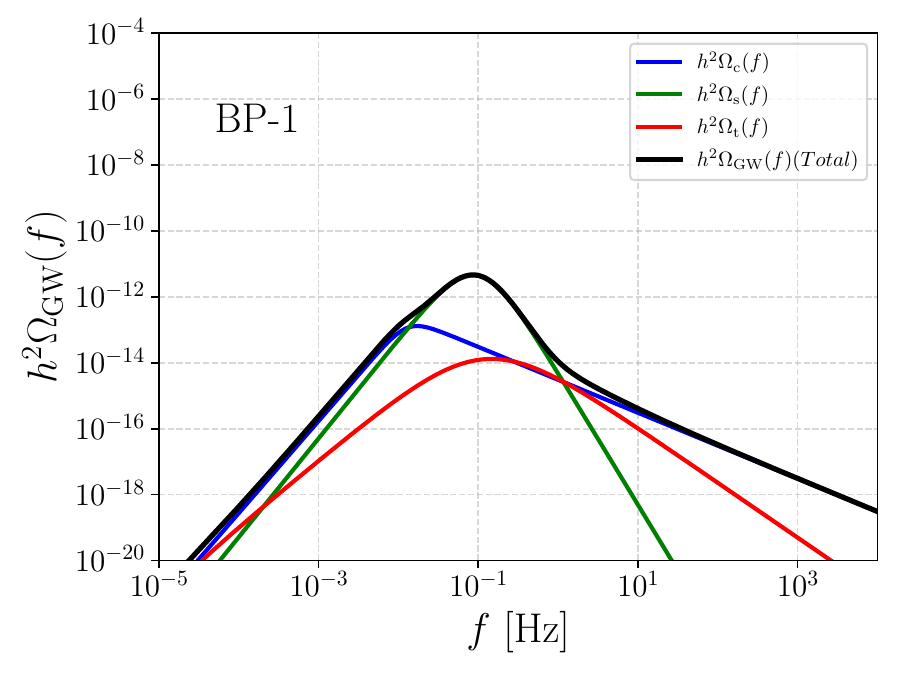}
\caption{Different contributions to the GW spectrum from a FOPT for representative values of the model parameters BP-1 listed in 
Table. \ref{tab:bp}. The peak amplitude of the spectrum is predominantly determined by the sound-wave contribution, whereas the spectral width is mainly controlled by the bubble-collision and magneto-hydrodynamics turbulence contributions.}
\label{fig:gw_all_contributions}
\end{figure}

The peak frequency of this MHD induced wave  is given by
\begin{equation}
    f_t = 2.7\times10^{-2}~\textrm{mHz} \left(\frac{g_*(T_*)}{100}\right)^{1/6}\left(\frac{T_*}{100~\GeV}\right)\left(\frac{\beta/H_*}{v_w}\right). \label{eq:fpeak_tur}
\end{equation}
We assume a non-runaway phase transition, for which the bubble-wall friction prevents indefinite acceleration of the wall and the bubble reaching a terminal velocity because the friction exerted by the surrounding plasma keeps overcoming the driving force from the vacuum energy. For a details discussion of limit on bubble wall velocity one can refer~\cite{Bodeker:2017cim}. In this regime, a significant fraction of the released energy is transferred to bulk fluid motion, making the sound-wave contribution the dominant component of the GW signal in the parameter region considered here. In this scenario the position and shape of the spectrum around the peak are primarily controlled by the acoustic contribution, whose characteristic spectral form and peak scale are supported by hydrodynamic simulations. Additional contributions from MHD turbulence and, depending on the bubble-wall dynamics, scalar-field bubble collisions can modify the spectrum away from the peak and contribute to the spectral tails~\cite{Hindmarsh:2013xza,Hindmarsh:2017gnf}. The different contributions of the stochastic GWs from FOPT for BP-1 are shown in Fig.~\ref{fig:gw_all_contributions}.
The sound-wave contribution $(\Omega_{s})$ is the most theoretically well known component of the GW spectrum, particularly with respect to its characteristic spectral shape, while treating its normalization and source lifetime with caution. The MHD turbulent contribution should generally be regarded to be less certain, especially in its amplitude and spectral form~\cite{Caprini:2019pxz}.\footnote{These theoretical uncertainties should be distinguished from uncertainties in the underlying phase-transition parameters, such as $\alpha$, $\beta/H_*$, $T_n$, and the bubble-wall velocity.}

\begin{figure}[H]
\centering
\includegraphics[width=0.7\linewidth]{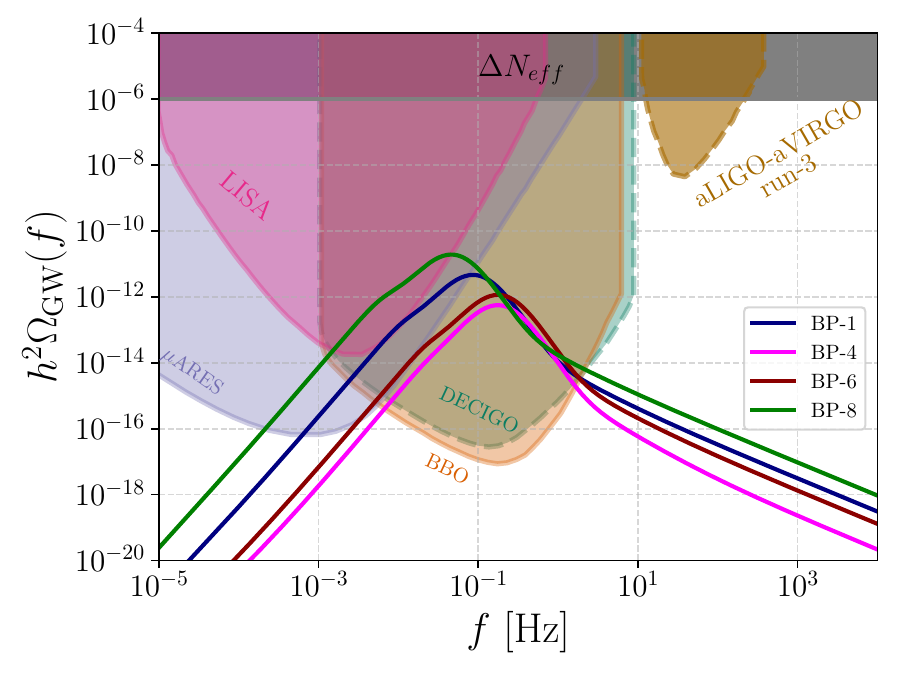}
\caption{The stochastic GW spectra predicted by our model for the selected benchmark points listed in Table~\ref{tab:bp}, together with the projected sensitivities of several future GW observatories. Also shown are the current constraints from Advanced LIGO and Advanced Virgo during the O3 observing run~\cite{KAGRA:2021kbb}. The rectangular grey-shaded region above $h^2\Omega_{\rm GW}>10^{-6}$ is excluded by constraints on $\Delta N_{\rm eff}$.}
\label{fig:gw}
\end{figure}

When the sound-wave contribution dominates the GW signal, the position of the peak  of the resultant GW spectrum is expected to be comparatively less uncertain than the spectral width and tails. The peak frequency is primarily determined by the characteristic length scale of the acoustic source, whereas the shape of the spectrum away from the peak can be modified by the less well-constrained contributions from MHD turbulence and scalar-field bubble collisions~\cite{Hindmarsh:2013xza,Hindmarsh:2017gnf,Caprini:2019pxz}.

There are several ongoing and proposed GW detection experiments covering a wide band of frequencies (in the nHz to kHz range). The list includes SKA~\cite{Weltman:2018zrl}, GAIA/THEIA~\cite{Garcia-Bellido:2021zgu}, AEDGE~\cite{AEDGE:2019nxb}, $\mu$ARES~\cite{Sesana:2019vho}, LISA~\cite{LISA:2017pwj}, DECIGO~\cite{Seto:2001qf}, ET~\cite{Punturo:2010zz}, and CE~\cite{Reitze:2019iox}, as well as recent proposals for high-frequency GW searches in the MHz--GHz range~\cite{Aggarwal:2020olq,Berlin:2021txa,Herman:2022fau,Bringmann:2023gba}. The GW spectra corresponding to four (4) benchmark points are presented in Fig.~\ref{fig:gw}. The calculated spectra in 4 cases, are superimposed on the region of sensitivity of 4 upciming experiments namely, LISA, $\mu$ARES, DECIGO as well as BBO. For comparison, we also show  the constraint on the GW spectrum from  the third observing run of advanced LIGO and advanced Virgo. The constraint on $\Delta N_{\rm eff}\,(\leq 0.18)$ from a joint BBN + CMB analysis~\cite{Yeh:2022heq}, translates into  an upper bound on GW energy density $h^2 \Omega_{\rm GW} \leq 5.6 \times 10^{-6} \Delta N_{eff}$~\cite{Caprini:2018mtu}. Corresponding region of exclusion from the consideration of $\Delta N_{\rm eff}$ is also shown in the figure by grey shaded region.
\begin{figure}[H]
\centering
\includegraphics[width=0.75\linewidth]{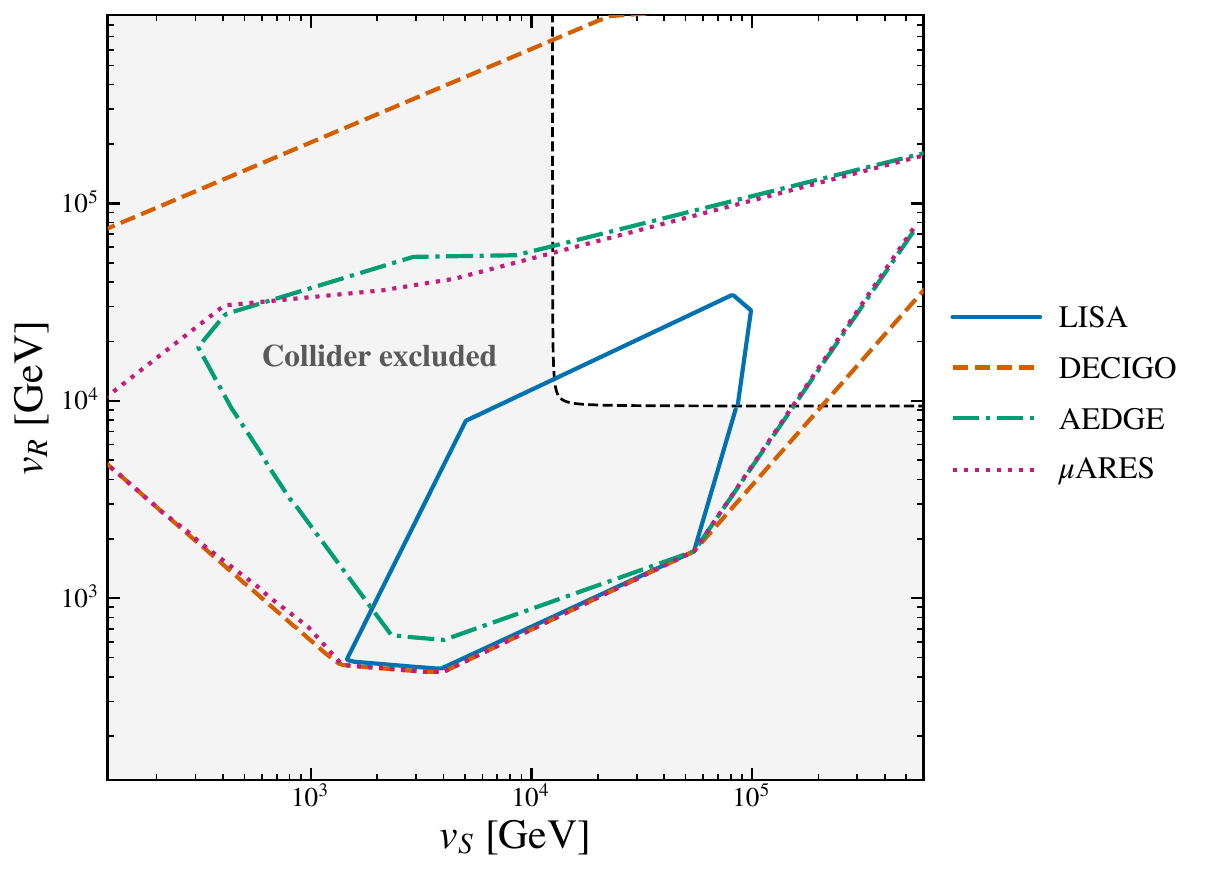}
\caption{Discovery sensitivity in the $(v_R-v_S)$ plane for an SNR $>10$ for various future GW observatories, with all other potential parameters fixed to their values at the benchmark point BP-1. The gray-shaded region indicates the collider-excluded parameter space, based on the lower mass bounds on the heavy charged and neutral gauge bosons.}
\label{fig:snr}
\end{figure}

A signal-to-noise ratio (SNR), $\rho$ can be defined to quantify the detectability of the stochastic GW signal by current and future GW observatories. $\rho$ is computed for each experiment using its corresponding sensitivity range and the knowledge of background noise for a given observation time $t_{\rm obs}$. $[f_{\min},f_{\max}]$~\cite{Allen:1996vm,Allen:1997ad,Maggiore:1999vm,Thrane:2013oya,Caprini:2019pxz}:
\begin{equation}
\rho = \left[n_{\rm det}{\rm t}_{\rm obs}\,\int_{f_\text{min}}^{f_\text{max}}\,\dd f\,\left(\frac{\Omega_\text{GW}(f)\,h^2}{\Omega_\text{noise}(f)\,h^2}\right)^2\right]^{1/2}\, . 
\label{eq:snr}
\end{equation}
$\Omega_{\rm noise}$ represents the detector noise. Different GW detectors determine their noise spectrum from the physical noise sources such as seismic noise, thermal noise, or pulsar timing noise and from the design of the detector. The resulting noise power spectral density is frequency dependent
Simulated noise data for each detector~\cite{Schmitz:2020syl,Sesana:2019vho,AEDGE:2019nxb}
available in a tabulated form, have been used to calculate the SNR.

$n_{\rm det}$ distinguishes between experiments employing auto-correlation ($n_{\rm det}=1$) and cross-correlation ($n_{\rm det}=2$) detection strategies. Throughout our numerical analysis, we assume $n_{\rm det}=1$ and adopt an observation time of ${\rm t}_{\rm obs}=1$ year for each experiment. A GW signal is considered detectable if the corresponding SNR satisfies $\rho>\rho_{\rm th}$, where $\rho_{\rm th}$ denotes the chosen detection threshold.

In Fig.~\ref{fig:snr}, the region bounded by the colored lines correspond to $\mathrm{SNR}>10$ for different experiments in the $v_R$–$v_S$ plane for the scalar potential parameters corresponding to BP-1.
It is evident from the plot that the GW observatories are capable probing any new physics 
beyond the SM as pointed out in the introduction that GW observations are sensitive to any physics beyond the SM operative at an energy scale, which is beyond the kinematic reach of even the LHC.  The DECIGO sensitivity lies well beyond an energy scale of $10^3 ~\rm TeV$. 
\section{Summary and Conclusion}\label{sec:con}
To summarise, we have investigated the dynamics of a first order phase transition and production of gravitational waves in the framework of a left right symmetric gauge theory. The scalar potential consists of a bi-doublet (under $SU(2)_L \otimes SU(2)_R$), a left handed doublet ($\Phi_L$), a right handed doublet $(\Phi_R)$ and a singlet Higgs fields ($\Phi_S$).  To begin with a Lagrangian which is invariant under $SU(3)_C \otimes SU(2)_L \otimes SU(2)_R \otimes U(1)_R \otimes U(1)_L$, the symmetry is broken spontaneously down to $SU(3)_C \otimes SU(2)_L \otimes U(1)_Y$ of Standard Model by $\langle\Phi_S\rangle$ and $\langle\Phi_R\rangle$, the non-zero vacuum expectation values  of scalar fields at an energy scale of $14 ~\rm TeV$ or higher.  During such epoch, when the Universe transited from a more symmetric vacuum configuration to a less symmetric vacuum, there is a possibility of first order phase transition taking place. Sufficiently strong first order phase transition in early Universe could have been the source of stochastic gravitational waves detected by several Gravitational wave observatories in recent time. 

We have investigated the nature and dynamics of the cosmological phase transition triggered by the Left-right symmetry breakdown 
at an energy scale of 10 TeV or higher. Due to the large hierarchy between the electroweak symmetry-breaking scale and the Left-Right symmetry-breaking scale, the dynamics of the Left-Right symmetry-breaking phase transition are primarily governed by the right-handed doublet $\Phi_R$ and the singlet $\Phi_S$. Here we would like to point out that we have restricted ourselves to the high scale phase transition dictated by the first breaking mentioned above. The bi-doublet $\Phi_B$ and the left-handed doublet $\Phi_L$ do not directly control the dynamics of this phase transition, although they contribute through loop and thermal corrections to the effective potential.

We have performed a detailed numerical analysis of the phase transition dynamics over a broad region of the model parameter space taking into account the possible theoretical and experimental constraints on the scalar potential parameters. We find that the model frequently exhibits FOPTs across a substantial portion of the parameter space. The resulting GW signals are generally weak and are unlikely to be detectable by current ground-based observatories such as LIGO. However, a significant fraction of the parameter space can produce GW signals within the projected sensitivity ranges of future space-based GW observatories, including LISA, $\mu$ARES, BBO, and DECIGO. Detection of such stochastic GW signals in the future detectors corroborates indirect evidence of any dynamics beyond the Standard Model operative at an energy scale of thousands of TeVs or higher well beyond the reach of present and proposed collider experiments.
\section*{Acknowledgment}
ARS acknowledges financial support from the Ministry of Minority Affairs, Government of India, through the Maulana Azad National Fellowship (F. No. 82-27/2019 (SA-III)). ND would like to acknowledge financial support from the ANRF grant CRG/2023/008234. AD acknowledges the hospitality of ARS and RA during his brief occasional visits at CTP, JMI.
\clearpage
\appendix
\section*{Appendix}
\section{Physical Spectrum and Mixing}
\label{app:physical_masses}
In this appendix, we present the masses and mixing parameters of all the physical scalar and gauge bosons in our model. In our analysis, we set all the new fermion Yukawa couplings to unity and take the SM fermion Yukawa couplings to their corresponding SM values. The masses of all the fermions can then be obtained from Eq.~\eqref{eq:lag_yukawa}. Therefore, we do not list the fermion masses and mixing parameters separately here.
\subsection{CP-even neutral scalars}
The CP-even mass matrix in this model involves mixing only among the $h_1^0$, $h_R^0$, and $h_S^0$ states. The other two states do not mix with any other states~\cite{Bhattacharyya:2021lgr}. The masses of these two states are given by
\begin{eqnarray}
    m_{h_2^{0}}^2 &=& \frac{1}{2} [4 \lambda_3 v_1^2 + (c_{14} -c_{13}) v_R^2] \nonumber \\
    m_{h_L^{0}}^2  &=& \frac{1}{2} (\rho_3 - 2 \rho_1) v_R^2 + \frac{1}{2}(c_{14} - c_{13}) v_1^2 ~.
\end{eqnarray}
The remaining three CP-even scalar states mix with each other. The corresponding effective $3\times3$ CP-even scalar mass matrix in the $\{h_1^0,h_R^0,h_S^0\}$ basis is given by
\begin{equation}
(M_{even}^2)_{3\times 3} =
\begin{pmatrix}
2\lambda_1 v_1^2 &
c_{13}v_1v_R &
\beta_1v_1v_S
\\[2mm]
c_{13}v_1v_R &
2\rho_1v_R^2 &
\gamma_1v_Rv_S
\\[2mm]
\beta_1v_1v_S &
\gamma_1v_Rv_S &
2\alpha_1v_S^2
\end{pmatrix}
\end{equation}
As $v_1 \ll v_R \sim v_S$, the mass matrix can be written in block form as
\begin{equation}
(M_{even}^2)_{3\times 3} =
\begin{pmatrix}
A & B^T\\
B & H
\end{pmatrix},
\end{equation}
where
\begin{equation}
A=2\lambda_1v_1^2,
\qquad
B=
\begin{pmatrix}
c_{13}v_1v_R\\
\beta_1v_1v_S
\end{pmatrix},
\qq
H= 
\begin{pmatrix}
2\rho_1v_R^2 & \gamma_1v_Rv_S\\
\gamma_1v_Rv_S & 2\alpha_1v_S^2
\end{pmatrix}.
\end{equation}
Since the two eigenvalues associated with the block $H$ are of order $v_R^2$ and $v_S^2$, while the light eigenvalue is of order $v_1^2$, the heavy degrees of freedom can be integrated out perturbatively. The resulting effective mass of the SM-like Higgs state is then obtained from the Schur complement (see-saw-like approximation)~\cite{Flieger:2020lbg},
given by
\begin{equation}
m_{h_1^0}^2 \simeq A-B^T H^{-1}B = 2v_1^2
\left[ \lambda_1 - \frac{ \alpha_1c_{13}^2 -\gamma_1c_{13}\beta_1 +\rho_1\beta_1^2}{4\alpha_1\rho_1 \gamma_1^2} \right].
\end{equation}
The remaining two eigenvalues are obtained by diagonalizing the
heavy-sector mass matrix $H$ and are given by
\begin{equation}
m_{h_R^0/ h_S^0}^2 \simeq \rho_1v_R^2+\alpha_1v_S^2 \pm
\sqrt{ \left(\rho_1v_R^2-\alpha_1v_S^2\right)^2 +\gamma_1^2v_R^2v_S^2 }.
\end{equation}
The mixing angle between the two heavy states is determined by
\begin{equation}
\tan 2\theta_H =
\frac{\gamma_1v_Rv_S}{\rho_1v_R^2-\alpha_1v_S^2}.
\end{equation}
The mixing angles of the light state with the two heavy states, $h_R$ and $h_S$, are respectively given by
\begin{align}
\theta_R &\simeq
\frac{v_1}{v_R} \frac{ 2\alpha_1c_{13}-\gamma_1\beta_1}{4\alpha_1\rho_1-\gamma_1^2}, \nonumber \\
\theta_S
&\simeq\frac{v_1}{v_S}
\frac{2\rho_1\beta_1-\gamma_1c_{13}}{4\alpha_1\rho_1-\gamma_1^2}.
\end{align}
\subsection{CP-odd neutral scalars}
The non-zero masses of the CP-odd neutral scalars are given by
\begin{eqnarray}
    m_{\xi_1^0}^2 &=& m_{\xi_R^{0}}^2 = m_{\xi_S^{0}}^2  = 0 \nonumber \\
    m_{\xi_2^{0}}^2 &=& \frac{1}{2} [4 \lambda_3 v_1^2 + (c_{14} -c_{13}) v_R^2] \nonumber \\
    m_{\xi_L^{0}}^2  &=& \frac{1}{2} (\rho_3 - 2 \rho_1) v_R^2 + \frac{1}{2}(c_{14} - c_{13}) v_1^2 
\end{eqnarray}
\subsection{Charged scalars}
The non-zero masses of the charged scalars are given by
\begin{eqnarray}
     m^2_{H_1^{\pm}} &=& \frac{1}{2} (c_{14}-c_{13}) (v_1^2 + v_R^2) \nonumber \\
     m^2_{H_L^{\pm}} &=&  \frac{1}{2} (\rho_3 - 2 \rho_1) v_R^2 \nonumber \\
     m_{H_2^{\pm}} &=&   m_{H_R ^{\pm}} =0
\end{eqnarray}
\subsection{Gauge bosons}
The squared masses of the charged gauge bosons are given by
\begin{equation}
	M_{W_L^{\pm}}^2 = \frac{1}{4} g_{2}^2 v_1^2 \qquad
    M_{W_R^{\pm}}^2 = \frac{1}{4} g_{2}^2 (v_1^2 + v_R^2)
\end{equation}
The squared mass matrix of the neutral gauge bosons in the
$\{W_{3L}, W_{3R}, B_L, B_R\}$ basis is given by 
\begin{eqnarray}
	M_{NG}^2 =  
    \begin{pmatrix}
	\frac{g_{2}^2 v_1^2}{4} & -\frac{g_{2}^2 v_1^2}{4} & \frac{g_{1} g_{2} v_1^2}{12} & -\frac{g_{1} g_{2} v_1^2}{12} \\ 
    
	-\frac{g_{2}^2 v_1^2}{4}  & \frac{g_{2}^2 (v_1^2+v_R^2)}{4} & -\frac{g_{1} g_{2} (v_1^2 + 2 v_R^2)}{12} & \frac{g_{1} g_{2} (v_1^2-v_R^2)}{12} \\
    
	\frac{g_{1} g_{2} v_1^2}{12} & -\frac{g_{1} g_{2} (v_1^2 + 2 v_R^2)}{12} & \frac{g_{1}^2 (v_1^2 + 4 v_R^2 + 4 v_S^2)}{36} & \frac{g_{1}^2(- v_1^2 + 2 v_R^2- 4 v_S^2)}{36} \\
    
	-\frac{g_{1} g_{2} v_1^2}{12} & \frac{g_{1}g_{2} (v_1^2-v_R^2)}{12} & \frac{g_{1}^2 (-v_1^2 + 2 v_R^2 - 4 v_S^2)}{36} & \frac{g_{1}^2 (v_1^2+ v_R^2 + 4 v_S^2)}{36} 
	\end{pmatrix}
\end{eqnarray}
Since this matrix has a zero determinant, at least one of its eigenvalues is zero, corresponding to the massless physical photon.
\section{Field-dependent Masses and Counterterms Coefficients}
\label{app:field_dependent_masses}
In this appendix, we collect the field-dependent squared mass matrices entering the one-loop effective potential. 
\subsection{CP-even neutral scalars}
The field-dependent squared mass matrix of the CP-even neutral scalars is
\begin{equation}
\mathcal{M}_{\rm E}^2 = \frac12
\scriptsize
\begin{pmatrix}
-2\mu_1^2+c_{13}\phi_R^2+\beta_1\phi_S^2 & 0 & 0 & 0 & 0\\
0 & -2\mu_1^2+c_{14}\phi_R^2+\beta_1\phi_S^2 & 0 & 0 & 0\\
0 & 0 & -2\mu_3^2+\rho_3\phi_R^2+\gamma_1\phi_S^2 & 0 & 0\\
0 & 0 & 0 & -2\mu_3^2+6\rho_1\phi_R^2+\gamma_1\phi_S^2 & 2\gamma_1\phi_R\phi_S\\
0 & 0 & 0 & 2\gamma_1\phi_R\phi_S & -2\mu_4^2+\gamma_1\phi_R^2+6\alpha_1\phi_S^2
\end{pmatrix}.
\end{equation}
\subsection{CP-odd neutral scalars}
The field-dependent squared mass matrix of the CP-odd neutral scalars is diagonal and may be written as
\begin{equation}
\mathcal{M}_{\rm O}^2
= \frac12
\scriptsize
\begin{pmatrix}
-2\mu_1^2+c_{13}\phi_R^2+\beta_1\phi_S^2 & 0 & 0 & 0 & 0\\
0 & -2\mu_1^2+c_{14}\phi_R^2+\beta_1\phi_S^2 & 0 & 0 & 0\\
0 & 0 & -2\mu_3^2+\rho_3\phi_R^2+\gamma_1\phi_S^2 & 0 & 0\\
0 & 0 & 0 & -2\mu_3^2+2\rho_1\phi_R^2+\gamma_1\phi_S^2 & 0\\
0 & 0 & 0 & 0 & -2\mu_4^2+\gamma_1\phi_R^2+2\alpha_1\phi_S^2
\end{pmatrix}.
\end{equation}
\subsection{Charged scalars}
The field-dependent squared mass matrix of the charged scalars is also diagonal,

\begin{equation}
\mathcal{M}_{\rm C}^2
= \frac12
\scriptsize
\begin{pmatrix}
-2\mu_1^2+c_{13}\phi_R^2+\beta_1\phi_S^2 & 0 & 0 & 0\\
0 & -2\mu_1^2+c_{14}\phi_R^2+\beta_1\phi_S^2 & 0 & 0\\
0 & 0 & -2\mu_3^2+\rho_3\phi_R^2+\gamma_1\phi_S^2 & 0\\
0 & 0 & 0 & -2\mu_3^2+2\rho_1\phi_R^2+\gamma_1\phi_S^2
\end{pmatrix}.
\end{equation}
\subsection{Gauge bosons}
The field-dependent squared masses of the charged gauge bosons and the field-dependent squared mass matrix of the neutral gauge bosons are
\begin{equation}
\mathcal{M}_{W_R^\pm}^2
= \frac{g_2^2}{4}\phi_R^2,
\end{equation}
\begin{equation}
\mathcal{M}_{NG}^2=
\begin{pmatrix}
\dfrac{g_2^2}{4}\phi_R^2 &
-\dfrac{g_2g_1}{6}\phi_R^2 &
-\dfrac{g_2g_1}{12}\phi_R^2\\[6pt]
-\dfrac{g_2g_1}{6}\phi_R^2 &
\dfrac{g_1^2}{9}(\phi_R^2+\phi_S^2) &
\dfrac{g_1^2}{18}(\phi_R^2-2\phi_S^2)\\[6pt]
-\dfrac{g_2g_1}{12}\phi_R^2 &
\dfrac{g_1^2}{18}(\phi_R^2-2\phi_S^2) &
\dfrac{g_1^2}{36}(\phi_R^2+4\phi_S^2)
\end{pmatrix}.
\end{equation}
\subsection{Fermions}
The field-dependent squared masses of the fermions are
\begin{equation}
\mathcal{M}_{q_S}^2 = \frac{y_s^2}{2}\phi_S^2,
\qquad
\mathcal{M}_{E}^2 = \mathcal{M}_{N}^2 = \frac{y_{LB}^2}{2}\phi_S^2.
\end{equation}
\subsection{Counterterms}
\begin{equation}
\begin{aligned}
\delta\mu_3^2 &=
\frac{3\Delta_R}{2v_R}
-\frac{\Delta_{RR}}{2}
-\frac{\Delta_{RS}v_S}{2v_R}, \\
\delta\mu_4^2 &=
\frac{3\Delta_S}{2v_S}
-\frac{\Delta_{SS}}{2}
-\frac{\Delta_{RS}v_R}{2v_S}, \\
\delta\rho_1 &=
\frac{\Delta_R}{2v_R^3}
-\frac{\Delta_{RR}}{2v_R^2}, \\
\delta\alpha_1 &=
\frac{\Delta_S}{2v_S^3}
-\frac{\Delta_{SS}}{2v_S^2}, \\
\delta\gamma_1 &=
-\frac{\Delta_{RS}}{v_Rv_S}.
\end{aligned}
\label{eq:counterterms}
\end{equation}
where
$\Delta_i \equiv 
\left.\frac{\partial V_{\mathrm{CW}}}{\partial \phi_i}\right|_{\phi_R=v_R,\phi_S=v_S}$  and 
$\Delta_{ij} \equiv 
\left.\frac{\partial^2 V_{\mathrm{CW}}}{\partial \phi_i \partial \phi_j}\right|_{\phi_R=v_R,\phi_S=v_S}$ with $i,j = R,S$.
\bibliographystyle{JHEP}
\bibliography{ref}
\end{document}